\documentclass[journal]{IEEEtran}
\usepackage{amsmath,amsfonts}
\usepackage{algorithm}
\usepackage{array}
\usepackage{mathrsfs}
\usepackage{textcomp}
\usepackage{stfloats}
\usepackage{url}
\usepackage{verbatim}
\usepackage{graphicx}
\usepackage{cite}
\usepackage{algpseudocode}
\usepackage{mathrsfs}
\usepackage{xcolor,soul}
\usepackage{adjustbox}
\usepackage{caption}
\usepackage{subcaption}

\newcommand{\reviewone}[1]{\textcolor{teal}{#1}}    %

\usepackage[normalem]{ulem}
\newcommand{\remove}[1]{\textcolor{blue}{\ifmmode\text{\sout{\ensuremath{#1}}}\else\sout{#1}\fi}}  
 
\usepackage{titlesec}
 \titlespacing*{\subsection}
{0pt}{ .1ex plus .1ex minus .2ex}{ .3ex plus .2ex}
\begin{document}

\title{Bayes2IMC: In-Memory Computing for  Bayesian Binary Neural Networks}

\author{Prabodh Katti,~\IEEEmembership{Student Member,~IEEE,} Clement Ruah, \IEEEmembership{Student Member,~IEEE,} Osvaldo Simeone, \IEEEmembership{Fellow,~IEEE,} Bashir M. Al-Hashimi, \IEEEmembership{Fellow, ~IEEE,} Bipin Rajendran, \IEEEmembership{Senior Member,~IEEE}\\
\thanks{The authors are affiliated with the Centre for Intelligent Information Processing Systems (CIIPS), Department of Engineering, King's College London, London WC2R 2LS, U.K. (bipin.rajendran@kcl.ac.uk)}
}



\maketitle

\begin{abstract}
Bayesian Neural Networks (BNNs) generate an ensemble of possible models by treating model weights as random variables. This enables them to provide superior estimates of decision uncertainty. However, implementing Bayesian inference in hardware is resource-intensive, as it requires noise sources to generate the desired model weights. In this work, we introduce Bayes2IMC, an in-memory computing (IMC) architecture designed for binary BNNs that leverages the stochasticity inherent to nanoscale devices. Our novel design, based on Phase-Change Memory (PCM) crossbar arrays eliminates the necessity for Analog-to-Digital Converter (ADC) within the array, significantly improving power and area efficiency. Hardware-software co-optimized corrections are introduced to reduce device-induced accuracy variations across deployments on hardware, as well as to mitigate the effect of conductance drift of PCM devices. We validate the effectiveness of our approach on the CIFAR-10 dataset with a VGGBinaryConnect model containing 14 million parameters, achieving accuracy metrics comparable to ideal software implementations. We also present a complete core architecture, and compare its projected power, performance, and area efficiency against an equivalent SRAM baseline, showing a $3.8$ to $9.6 \times$ improvement in total efficiency (in GOPS/W/mm$^2$) and a $2.2 $ to $5.6 \times$ improvement in power efficiency (in GOPS/W). In addition, the projected hardware performance of Bayes2IMC surpasses  most  memristive BNN architectures  reported in the literature,  achieving up to $20\%$ higher power efficiency compared to the state-of-the-art.
\end{abstract}


\section{Introduction}
\label{sec:Intro}
\subsection{Context and Motivation}
\begin{figure}[h]
    \centering
        \fbox{\includegraphics[width=0.325\textwidth]{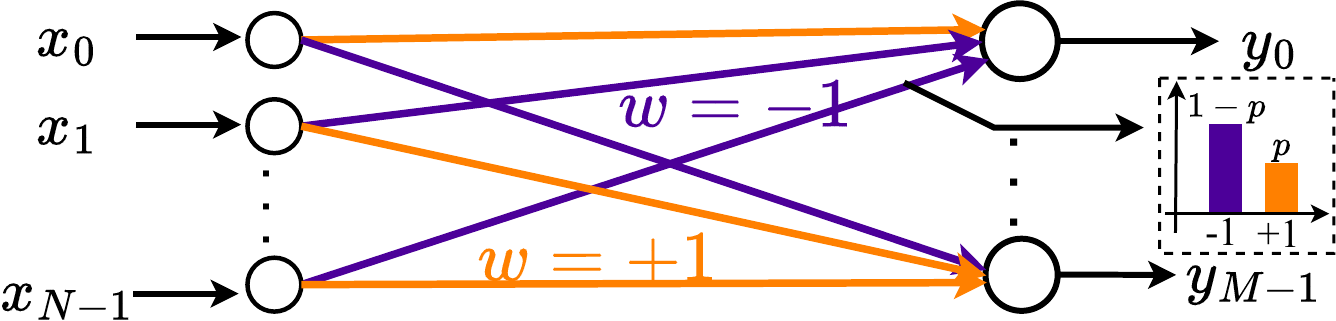}}
        
        \vspace{1mm}
        \fbox{\includegraphics[width=0.325\textwidth]{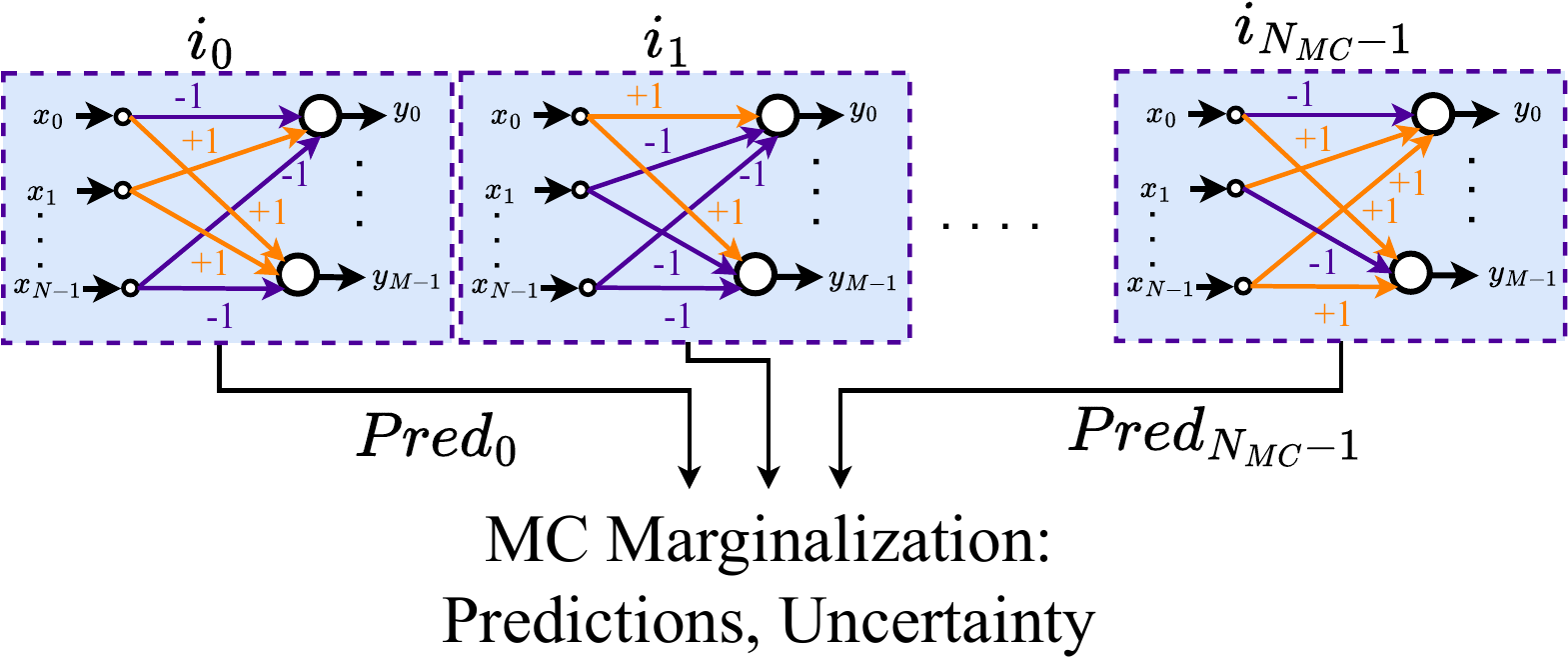}}
   
    \vspace{1mm}
        \fbox{\includegraphics[width=0.325\textwidth]{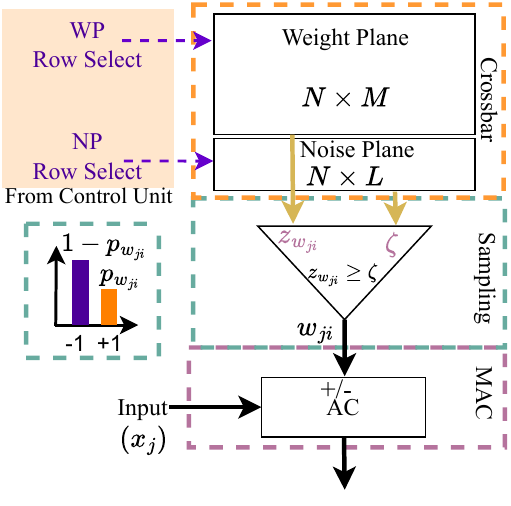}}
    \caption{\textbf{Top:} Illustration of a Bayesian neural network (BNN) where weights take binary values upon sampling. \textbf{Middle:} Bayesian inference is performed by an ensemble of  $N_{MC}$ predictions combined through Monte Carlo sampling to obtain predicted class, prediction confidence, and uncertainty. \textbf{Bottom:} Block diagram of the proposed Bayes2IMC core architecture implementing BNN inference. The crossbar array of memristive devices is divided into a weight plane (WP) and a noise plane (NP). The WP stores the parameters $z_{w_{ji}}$ obtained by reparametrizing the probability parameters $p_{w_{ji}}$,  and the noise plane generates the stochasticity required for synaptic sampling.  Binary weights $w_{ji}$ are then generated by comparing these variables in hardware. Unlike traditional IMC architectures, the input $x_j$, $j^{th}$ element of input vector $\mathbf{x}$, is accumulated based on the sign of $w_{ji}$. 
    }
    \label{fig:First}
    \vspace{-15pt}
\end{figure}

\IEEEPARstart{T}{he} growing demand for edge AI in applications such as medical diagnostics \cite{band2022benchmarking}, facial identification and surveillance \cite{Kendall_Gal_2017} and self-driving cars \cite{tambon2022certify} where reliability and safety are of paramount importance, has heightened the need for systems that provide a measure of uncertainty while operating under significant resource constraints in terms of area and power. Advances in deep neural networks have ensured very high accuracy, but this has come at the cost of overconfidence and poor calibration \cite{guo2017calibration}. Specifically, such networks are incapable of distinguishing between aleatoric uncertainty (data uncertainty) and epistemic uncertainty (model uncertainty),\cite{huang2024calibrating}, making tasks such as  out-of-distribution (OOD) detection particularly challenging. 
Bayesian neural networks (BNNs) overcome this problem by allowing the weights to be random variables rather than point estimates, thus encoding uncertainty in the parameter distributions \cite{10181438, jang2021bisnn}. To perform inference, the network weights are sampled to create an ensemble of predictors, whose outputs are then combined to obtain data class prediction and uncertainty estimates  (Fig.~\ref{fig:First}).

The creation of ensembles can be done in time, i.e., instantiating each ensemble one after another on the same hardware \cite{katti2024bayesian}, or in space, i.e., instantiating all the ensembles simultaneously, using multiple copies of the hardware \cite{10181438}. Both methods are   resource-intensive compared to a traditional network whose parameters are point estimates and  require only a single instantiation. 

The ensembling-in-time approach reduces net decision-making throughput while optimizing for   power and total area. In contrast, the ensembling-in-space expends more area and power in order to deliver high net throughput. In addition, both these approaches require noise sources to generate random numbers. Thus, the deployment of Bayesian neural networks at the edge is challenging and requires careful hardware-software co-optimization \cite{lu2022algorithm, bonnet2023bringing}. 

In this paper, we design and validate an area and power-efficient ensembling-in-time  Bayesian binary neural network in-memory compute (IMC) architecture named Bayes2IMC that leverages the inherent stochasticity of nanoscale memory devices. Fig.~\ref{fig:First} provides a high-level description of this core. The nanoscale devices are arranged in a crossbar fashion, which is partitioned into two sections: a weight plane (WP) that stores the distribution parameters, and a noise plane (NP) that provides stochasticity necessary for sampling. This approach was first introduced in the conference version \cite{10181438} of this work. We also modify the traditional in-memory computing approach by routing inputs to an accumulator for multiply-and-accumulate (MAC) operations. This combination of routing strategy, row-by-row read flow, and on-the-fly sample generation during sensing enables inference without the need for an analog-to-digital converter, thereby improving area and power efficiency.

\subsection{Background and Related Work}
\noindent\emph{In-Memory Computing}: 
The IMC approach has been proposed to address the `von Neumann bottleneck' \cite{wulf1995hitting} plaguing traditional computing platforms \cite{jain2019neural,aguirre2024hardware, verma2023neuromorphic,jhang2021challenges,8401845}. In such systems, certain computational operations are performed in  memory without data movement, enabling high throughput at low area and power.
IMC implementations based on digital  Static Random Access Memories (SRAMs) \cite{mittal2021survey, peng2019dnn+,8401845, kneip2021impact, ma2024dcim} as well as those based on non-volatile memory (NVM) devices such as Phase Change Memory (PCMs),  Resistive RAMs (RRAMs), Spin-Transfer Torque RAMs (STT-RAMs), and Spin-Orbit Torque Magnetic RAMs (SOT-MRAMs) have been explored for implementing frequentist deep learning models \cite{yang2020all, verma2023neuromorphic, yu2021rram}. 
Unlike SRAM IMC implementations, NVM devices allow multi-level conductance for multi-bit parameter storage, making them area and power-efficient. 

However, NVM devices also pose several challenges for IMC MVM implementation. The high level of state-dependent programming \cite{joshi2020accurate} results in write error, and conductance drift results in instability of programmed weights \cite{joshi2020accurate, bonnet2023bringing}, severely impacting the classification accuracy. Additional sources of error include MVM errors due to IR drops on source-line due to high current in the all-parallel operation mode \cite{peng2019dnn+}, and non-linear dependence of device current on applied voltage \cite{chen2015technology}. Furthermore, Analog-to-Digital converters (ADCs) are necessary to convert the analog output of the MVM operation to multi-bit digital values, and this  impacts both the overall power consumption and area efficiency \cite{peng2019dnn+, jain2019neural}.

\noindent\emph{Related works}: 
We now discuss proposed IMC solutions in the literature that leverage the inherent stochasticity of NVM devices with the goal of implementing ensembling methods. An RRAM array-based Bayesian network with training implemented via Markov Chain Monte Carlo (MCMC) was demonstrated in \cite{dalgaty2021situ}, though the need for training in situ is time and resource intensive    \cite{salimans2015markov}. Bernoulli distributed noise of Domain Wall Magnetic Tunnel Junction (DW-MTJ) devices and SOT-MRAM were used in \cite{yang2020all} and \cite{lu2022algorithm}  respectively to generate Gaussian random variables. Since the weights are Gaussian distributed, two parameters -- the mean and standard deviation -- are required for each weight. In addition, generating a single Gaussian variable using the central limit theorem requires contributions from multiple noise sources, leading to expensive hardware overhead.


An efficient implementation was presented in \cite{bonnet2023bringing} that performs device-aware training and combines multiple devices to generate the required stochasticity per weight parameter. The authors of \cite{ahmed2023spindrop} use STT-RAM to implement binary neural networks and perform Bayesian inference using Monte-Carlo Dropout (MCD) methods \cite{pmlr-v48-gal16}, where neurons are randomly dropped out during inference leveraging a random number generated using STT-RAM's stochasticity. While MCD-based networks are easy to train and deploy, they are reported to be less expressive, especially for out-of-distribution uncertainty estimations \cite{9756596, pmlr-v119-chan20a}.

To mitigate effects of conductance drift, periodic calibrations of each layer of the network was used to obtain optimal affine factors to reduce error in \cite{joshi2020accurate}. Alternatively, \cite{bonnet2023bringing},  proposed reprogramming at different time intervals based on the effect of drift on the available domain of programmable conductance values. However, this involves multiple reprogramming steps, which is resource-intensive. 

Many of these works also employ a digital-to-analog-converter (DAC) to provide analog voltage input vectors. This can result in errors due to I-V non-linearity in some devices, as discussed in \cite{chen2015technology}. Working around this problem, references \cite{peng2019dnn+, lu2022algorithm,bonnet2023bringing} propose bit-slicing, wherein the input bits are sent one by one, and the accumulated outputs after analog-to-digital conversion are scaled by their binary positional weights and summed. 
Pulse width modulation (PWM) can also be used, as suggested in \cite{aguirre2024hardware}. 
In addition to extra circuitry overhead -- shift-and-add circuits in the former or PWM generator in the latter -- all the methods described require an ADC for readout, which consumes up to $87\%$ power and $60\%$ of the total area of a core \cite{jain2019neural,yu2021rram}. To reduce the power and area overhead, the ADC use is typically multiplexed amongst several columns, affecting the overall throughput \cite{peng2019dnn+}.

\subsection{Main Contributions}
In this work, we introduce Bayes2IMC, a PCM-based computing architecture designed for BNNs with binary weights. Binary BNNs are less resource-intensive because they need only one parameter to describe the weight distribution. 
 The main contributions of the paper are as follows:
\begin{itemize}
    \item We formulate a principled way to utilize the inherent device stochasticity as a noise source for ensembling.
    \item We devise a hardware-software co-optimized technique that is applied only on the output logits to reduce accuracy variations across multiple network deployments. 
    \item We propose a simple global drift compensation mechanism to reverse the effect of state-dependent conductance drift on network performance. This does not require frequent reprogramming or layer-by-layer calibration, only a simple rescaling of the read pulse. Our approach ensures no drop in accuracy, expected calibration error (ECE) \cite{guo2017calibration} and uncertainty performance up to $10^7$ seconds.
    \item We propose a crossbar array-based architecture that does not require complex ADCs. It employs a novel scheme that generates binary weights based on the simultaneous summation of conductances from the weight plane and noise plane. This permits synaptic weights to be generated on-the-fly in a power and area-efficient manner, enabling an ensembling-in-time implementation.
    \item  Unlike traditional IMC accelerators, our approach streams the inputs directly into the pre-neuron accumulator, avoiding expensive DACs in every row of the crossbar.  With this approach,  read pulses with fixed amplitude and pulse width can be applied on the bit-line for sampling, thus avoiding MVM errors due to the non-linear dependence of the device conductance on the applied voltage. This is markedly different from conventional IMC methods that require streaming the input bits of the activation variable bit-by-bit and performing shift-add operations to obtain MVM output \cite{7092504, bonnet2023bringing}.  
\end{itemize}
 To the best of our knowledge, this is the first work to demonstrate variational inference-based binary BNNs on IMC. We achieve a projected total efficiency improvement of $3.8-9.6 \times$ as compared to an equivalent SRAM architecture. 
 
 The rest of the paper is organized as follows. In Section \ref{Sec: Bkgrnd}, we review background material on Bayesian neural networks, uncertainty quantification and PCM devices. In Section \ref{sec:AnD}, we discuss the design, architecture and the co-optimization between algorithm or software and hardware. The accuracy performance and the hardware projections are discussed in Section \ref{sec:Results}. Finally, Section \ref{sec:Conclusion} concludes the paper.

\section{Preliminaries}
\label{Sec: Bkgrnd}

In this section, we review preliminary material on binary BNNs (BBNN), calibration measures, aleatoric and epistemic uncertainties and PCM devices.
\subsection{Binary Bayesian Neural Networks}

Bayesian neural networks (BNNs) predict an output $\boldsymbol{y}$ for a given input $\boldsymbol{x}$ by evaluating the posterior distribution $p(\boldsymbol{w}|\mathcal{D})$, where  the posterior distribution quantifies the likelihood of  $\boldsymbol{w}$ under the evidence presented by the training dataset $\mathcal{D} = \{x_i, y_i\}_i$. At inference time, the contribution of each plausible model $p(\boldsymbol{y}|\boldsymbol{x},\boldsymbol{w})$ is weighted by the posterior $p(\boldsymbol{w}|\mathcal{D})$, yielding the marginal distribution \cite{wilson2020case, 9756596}
\begin{equation}
    p(y|\boldsymbol{x},\mathcal{D})
    = \mathbb{E}_{p(\boldsymbol{w}|\mathcal{D})} \left[ p(y|\boldsymbol{x},\boldsymbol{w}) \right].
    \label{formalinf}
\end{equation}
In this work, we focus on the implementation of BBNNs, where each weight $w$ is restricted to a binary set of values $w \in \{-1, 1\}$.
Accordingly, the parameter space is defined as $\mathcal{W} = \{-1, 1\}^{\vert \boldsymbol{w} \vert}$, where $\vert \boldsymbol{w} \vert$ represents the number of synaptic weights.
Intuitively, BBNNs implement a set of neural networks where the weights are trained to follow a posterior distribution $p(\boldsymbol{w}|\mathcal{D})$ rather than a single point-estimate \cite{9756596}.
 BBNNs are then trained   through mean-field variational inference (VI) as discussed in \cite{khan2023bayesian, Meng_Bachmann_Khan_2020}, \reviewone{where the} posterior distribution $p(\boldsymbol{w}|\mathcal{D})$ is approximated by a parameterized distribution $q(\boldsymbol{w} | \boldsymbol{\lambda})$, where variational parameter $\boldsymbol{\lambda}$ consists of natural parameters of Bernoulli distribution $\lambda_w$ in the exponential family, i.e $\boldsymbol{\lambda} = \{\lambda_w\}_{w \in \boldsymbol{w}} \in \mathbb{R}^{\vert \boldsymbol{w} \vert}$. The probability parameter $p_w$ describing the distribution of binary weights $w$ is encoded as
\begin{equation}
    p_w = \frac{1}{1 + e^{-2\lambda_w}}.
    \label{pwlamdaw}
\end{equation}
$\boldsymbol{\lambda}$ is trained using the evidence lower bound (ELBO) criterion or free energy, which can be represented as \cite{Simeone_2022}
\begin{equation}
    \underset{\boldsymbol{\lambda}}{\mathrm{min}}\ \mathbb{E}_{q(\boldsymbol{w} | \boldsymbol{\lambda})}[\mathcal{L}_D(\boldsymbol{w})] + \mathrm{KL}(q(\boldsymbol{w} | \boldsymbol{\lambda})||p(\boldsymbol{w})),
    \label{eq:ELBO}
\end{equation}
where $\mathrm{KL(\cdot || \cdot)}$ is the KL divergence. 

Training is performed using the BiNN optimizer, which is described in \cite{Meng_Bachmann_Khan_2020}, with ELBO criterion \eqref{eq:ELBO} as the optimization problem to obtain a collection of synaptic probabilities $\boldsymbol{p}_{\boldsymbol{w}} = \{p_w\}_{w \in \boldsymbol{w}}$ from \eqref{pwlamdaw} (further details on training are provided in Supplementary Section A). Using an appropriate binary random variable generator, $\boldsymbol{p}_{\boldsymbol{w}}$ can be leveraged to create an ensemble of networks with $\boldsymbol{w} \sim q(\boldsymbol{w} | \boldsymbol{\lambda})$.
As illustrated in Fig.~\ref{fig:First}, during inference, the outputs $p(y|\boldsymbol{x},\boldsymbol{w})$ of the sampled networks are aggregated to obtain the final marginal predictor 

\begin{equation}
    p(y|\boldsymbol{x},\mathcal{D})
    \approx \frac{1}{N_{MC}}\sum_{i=1}^{N_{MC}} p(y|\boldsymbol{x},\boldsymbol{w}_i),
    \label{eq:ensembling}
\end{equation}
where $N_{MC}$ is the number of independent and identically sampled (i.i.d.) networks $\{\boldsymbol{w}_i\}_{i=1}^{N_{MC}} \overset{\mathrm{i.i.d.}}{\sim} q(\boldsymbol{w} | \boldsymbol{\lambda})$ \cite{9756596, Meng_Bachmann_Khan_2020}. We focus on classification tasks in this paper, where the target variable $y$ takes values in a discrete set $\mathscr{C} = \{0, 1, ..., n-1\}$ for $n \geq 1$. Accordingly, from the set confidence levels $\{ p(y|\boldsymbol{x},\mathcal{D}) \}_{y \in \mathscr{C}}$, a prediction is obtained as $\hat{y} =  \mathrm{argmax}_{y \in \mathscr{C}} \{ p(y|\boldsymbol{x},\mathcal{D}) \}$.

\begin{figure*}[htbp]
    \centering
\begin{minipage}[t]{0.24\textwidth}
        \centering
        \begin{minipage}[t]{\textwidth}
            \centering
            \includegraphics[width=\textwidth]{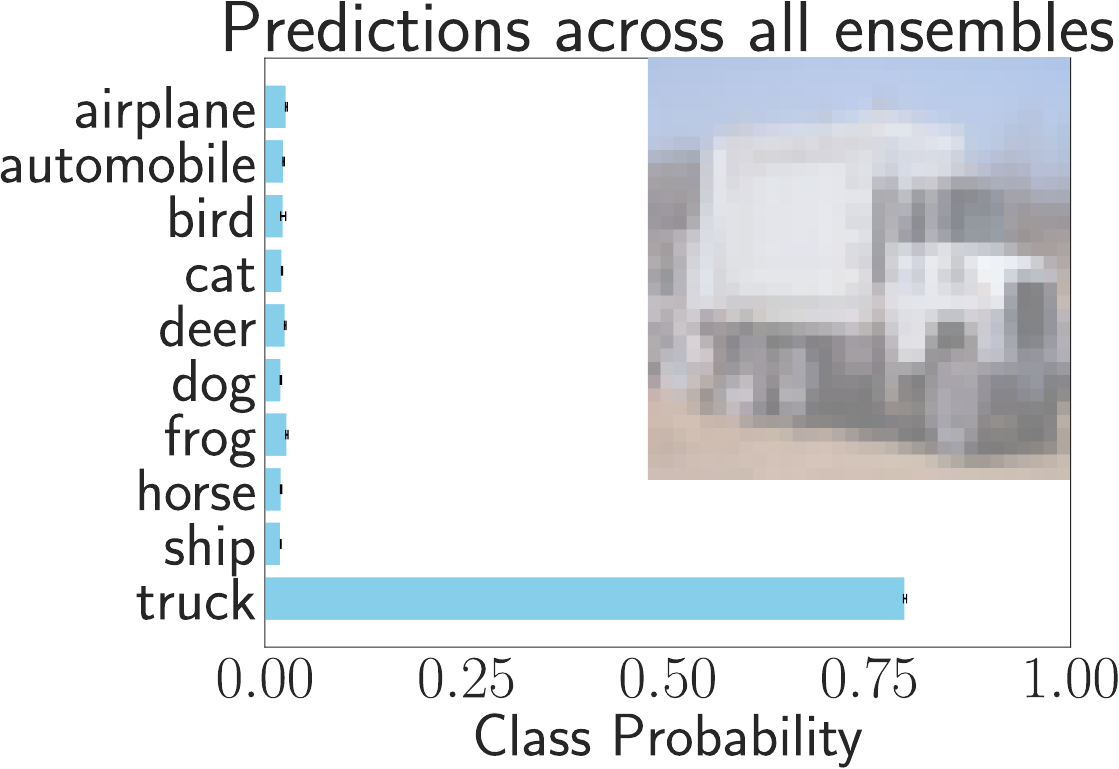}
        
        \end{minipage}
        \subcaption{Dataset: CIFAR10 (IND);  Prediction: Correct}
        \label{fig:subfig1}
    \end{minipage}
    \begin{minipage}[t]{0.24\textwidth}
        \centering
        \begin{minipage}[t]{\textwidth}
            \centering
            \includegraphics[width=\textwidth]{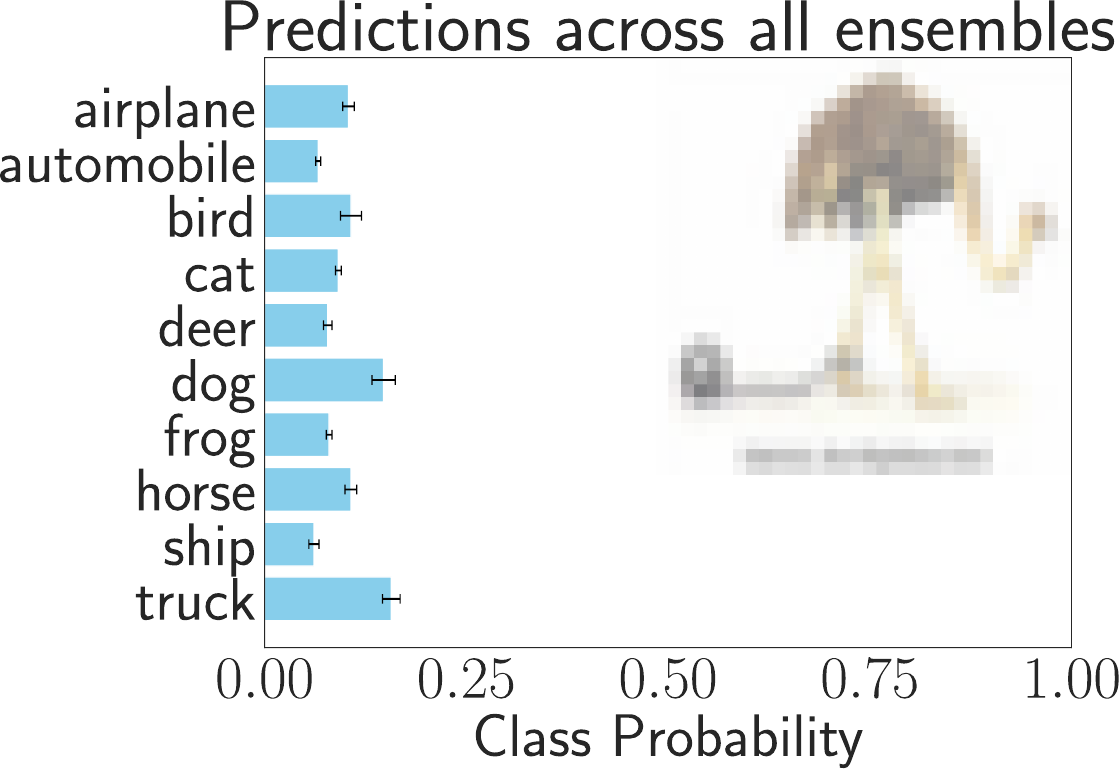}
        
        \end{minipage}
        \subcaption{ Dataset: CIFAR10 (IND); Prediction: Incorrect}
        \label{fig:subfig2}
    \end{minipage}
    \begin{minipage}[t]{0.245\textwidth}
        \centering
        \begin{minipage}[t]{\textwidth}
            \centering
            \includegraphics[width=\textwidth]{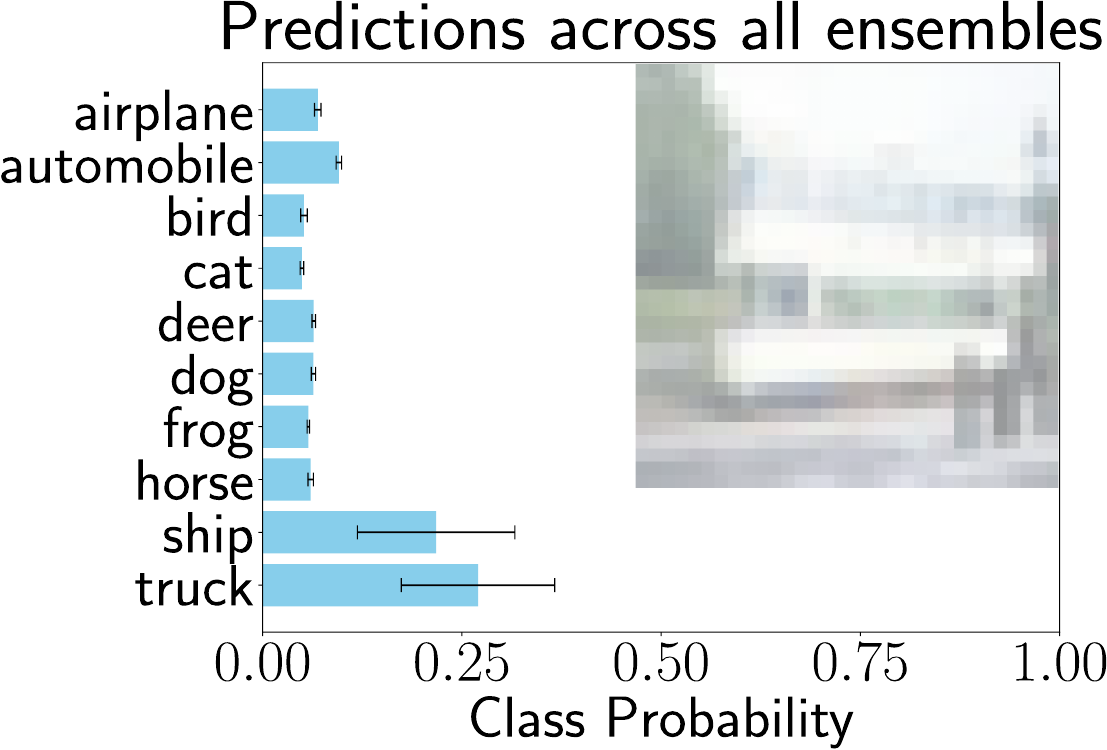}
    
        \end{minipage}
        \subcaption{ Dataset: CIFAR100 (OOD)}
        \label{fig:subfig3}
    \end{minipage}
    \begin{minipage}[t]{0.24\textwidth}
        \centering
        \begin{minipage}[t]{\textwidth}
            \centering
            \includegraphics[width=\textwidth]{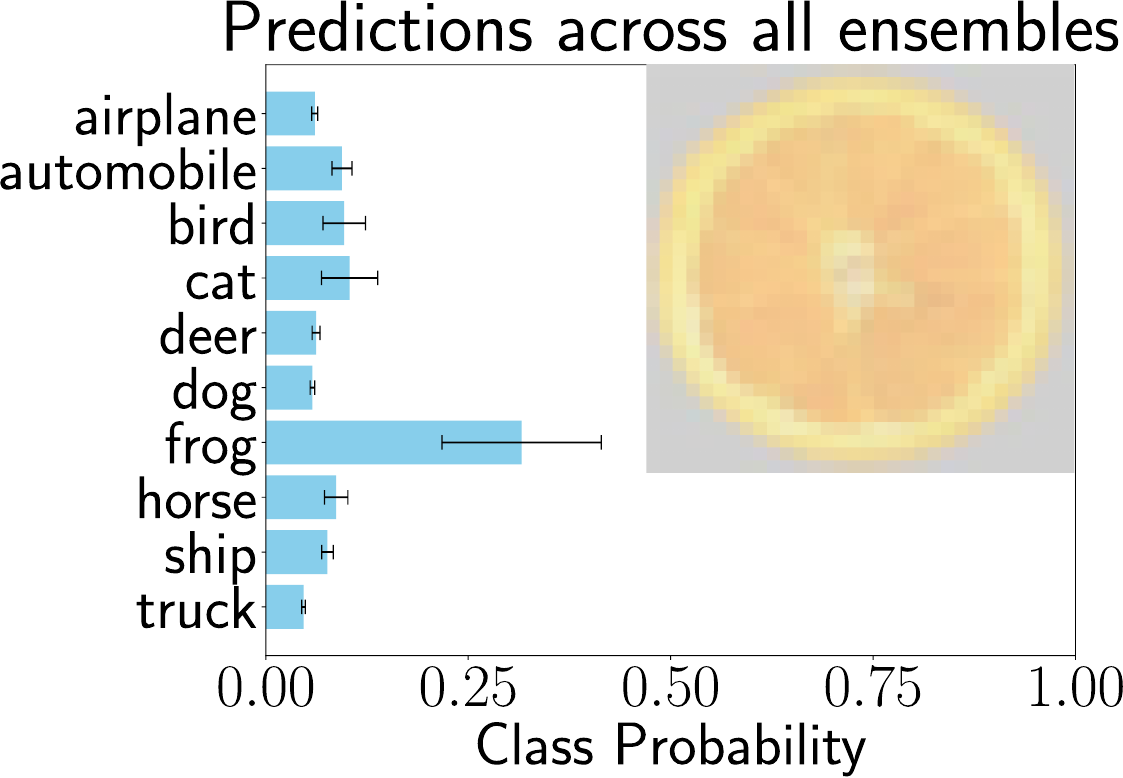}
        \end{minipage}
        \subcaption{ Dataset: CIFAR100 (OOD)}
        \label{fig:subfig4}
    \end{minipage}

   \caption{Illustration of aleatoric and epistemic uncertainty using sample images from the CIFAR10 and CIFAR100 datasets. The results shown are generated from the outputs of a neural network implemented with Bayes2IMC architecture on a custom PCM simulator. This network was trained on the CIFAR10 dataset. Given that CIFAR100 and CIFAR10 contain mutually exclusive classes, the CIFAR100 dataset was utilized to evaluate OOD detection performance. The error bars indicate the variability of probabilities, capturing disagreement amongst predictors in the ensemble. 
   }
    \label{fig:uncert}
    \vspace{-10pt}
\end{figure*}

\subsection{Expected calibration error (ECE)}
The ECE indicates the in-distribution calibration of a trained network by quantifying the expected discrepancy between accuracy and the confidence scores obtained by the output probability of the predicted class.  For a given dataset $\mathcal{D} = \{x_i, y_i\}_{i=1}^{N_{s}}$ of $N_s$ samples, we denote as $B_m = \{ (\hat{p}_i, \hat{y}_i, y_i) \}_{i | \hat{p}_i \in I_m}$ the set of confidence-label pairs for which the predicted confidence $\hat{p}_i = p(y_i | x_i, \mathcal{D})$ lies inside $I_m$, with $m \in \{1, ..., N_b\}$. Here $N_b$ denotes the number of equally spaced intervals in $[0,1]$ and $I_m = \big((m-1)/N_b,m/N_b\big]$.
The ECE is mathematically described, as per \cite{guo2017calibration}, as
\begin{equation}
    ECE = \sum_{m=1}^{N_b}\frac{|B_m|}{N_s}\bigg|\mathrm{acc}(B_m)-\mathrm{conf}(B_m)\bigg|,
    \label{eq:ECE}
\end{equation}
where $\mathrm{acc}(B_m) = \frac{1}{|B_m|} \sum_{(\hat{y}, y) \in B_m}\delta_{\{\hat{y} = y\}}$ is the number of correctly predicted samples in $B_m$ and $\mathrm{conf}(B_m) = \frac{1}{|B_m|} \sum_{\hat{p} \in B_m} \hat{p}$ is the average confidence or probability of the predicted class in the interval. 


\subsection{Aleatoric and Epistemic Uncertainty}

For a given covariate $\boldsymbol{x}$, the total uncertainty of the predictions can be quantified by computing the entropy of the ensemble predictor $p(y|\boldsymbol{x},\mathcal{D})$ \cite{bonnet2023bringing},
\begin{multline}
U_{tot}= \mathbb{E}_{y \sim p(y | \boldsymbol{x},\mathcal{D})} \left[ -\log\left( p(y | \boldsymbol{x},\mathcal{D}) \right) \right] \\
    \approx - \sum_{y \in \mathscr{C}} \frac{1}{N_{MC}}\sum_{i=1}^{N_{MC}} p(y|\boldsymbol{x},\boldsymbol{w}_i) \log\bigg(\frac{1}{N_{MC}}\sum_{i=1}^{N_{MC}} p(y|\boldsymbol{x},\boldsymbol{w}_{i})\bigg),\\
    \label{eq:up}
\end{multline}


where $\{\boldsymbol{w}_i\}_{i=1}^{N_{MC}} \overset{\mathrm{i.i.d.}}{\sim} q(\boldsymbol{w} | \boldsymbol{\lambda})$ and
$p(y|\boldsymbol{x},\boldsymbol{w}_i)$ is the probability value for the given class $y \in \mathscr{C}$.
This total uncertainty $U_{tot}$ can be decomposed into the uncertainty due to the inherently stochastic nature of the covariate-label pairs $(\boldsymbol{x}, y)$, i.e., the aleatoric uncertainty; and  the uncertainty stemming from the limited amount of evidence $\mathcal{D}$ i.e., the epistemic uncertainty.

Aleatoric uncertainty can be quantified by the entropy of the labels $y \sim p(y | \boldsymbol{x}, \boldsymbol{w})$ with average evaluated with respect to the random parameters $\boldsymbol{w} \sim q(\boldsymbol{w} | \boldsymbol{\lambda})$, which is given mathematically as
\begin{equation}
\begin{split}
U_a
    &= \mathbb{E}_{\boldsymbol{w} \sim q(\boldsymbol{w} | \boldsymbol{\lambda}), y \sim p(y | \boldsymbol{x},\boldsymbol{w})} \left[ -\log\left( p(y | \boldsymbol{x},\boldsymbol{w}) \right) \right] \\
    &\approx - \frac{1}{N_{MC}} \sum_{y \in \mathscr{C}} \sum_{i=1}^{N_{MC}}  p(y|\boldsymbol{x},\boldsymbol{w}_i) \log\left( p(y|\boldsymbol{x},\boldsymbol{w}_i) \right),
\end{split}
\label{eq:ua}
\end{equation}

for $\{\boldsymbol{w}_i\}_{i=1}^{N_{MC}} \overset{\mathrm{i.i.d.}}{\sim} q(\boldsymbol{w} | \boldsymbol{\lambda})$.
The epistemic uncertainty is evaluated instead as the difference between the total uncertainty and  aleatoric uncertainty, i.e.,
\begin{equation}
U_e = U_{tot}-U_a.
\label{eq:ue}
\end{equation}
Epistemic uncertainty is zero when the mean predictor $p(y | \boldsymbol{x},\mathcal{D})$ agrees with each individual predictor $p(y | \boldsymbol{x},\boldsymbol{w})$ of the ensemble, and is positive when they disagree. Note that for frequentist networks with fixed weights, $U_{tot}=U_a$, and therefore $U_e=0$.

Ideally, for an out-of-distribution (OOD) data, the ensemble of predictors will disagree on predictions, causing a large epistemic uncertainty. In contrast, on in-distribution (IND) data, there should be a broad agreement on the prediction probabilities, and thus the epistemic uncertainty should be low. Fig.~\ref{fig:uncert} illustrates the distinct roles that aleatoric and epistemic uncertainties play in understanding the predictions of a BBNN. 

Fig.~\ref{fig:subfig1} and Fig.~\ref{fig:subfig2} consider the output of the BBNN to IND data. In Fig.~\ref{fig:subfig1}, the models assign a relatively high probability score to one class (`Truck') and low scores for the rest, indicating low aleatoric uncertainty $(U_a)$. In Fig.~\ref{fig:subfig2}, the model assigns low and relatively similar probability scores to all classes, indicating higher $U_a$. The epistemic uncertainty ($U_e$) is low in both cases as there is agreement amongst the predictors on the outputs. This is indicated by error bars that signify the spread of probability scores per class across the ensemble.
Fig.~\ref{fig:subfig3} and Fig.~\ref{fig:subfig4} consider  OOD data corresponding to different types of inputs as compared to the trained data. In this case the high epistemic uncertainty is manifested by the disagreement amongst the probability distributions in the ensembles illustrated by a larger spread of probability values indicated by error bars due to the unfamiliar nature of the inputs.
\subsection{PCM}
Phase change memory (PCM) devices consist of chalcogenide materials such as Ge$_2$Sb$_2$Te$_5$, sandwiched between two metal electrodes.  
By adjusting the proportion and arrangement of the crystalline and amorphous phases of the chalcogenide material, PCM devices can be programmed to multiple analog non-volatile conductance states. \cite{doi:10.1063/1.5042408,rajendran2019building}. Programming a PCM device involves applying electric pulses, wherein the number, amplitude, and duration of these pulses are adjusted iteratively until the desired conductance is attained. The conductance thus programmed exhibits stochasticity that is observed to be state-dependent \cite{papandreou2011programming}.

\begin{figure}
  \centering
  
  \begin{minipage}{0.2\textwidth}
    \centering
    \includegraphics[width=\textwidth]{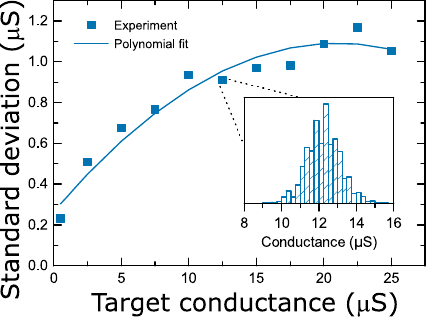}
    \subcaption{Programming noise.}
    \label{fig:Prog_Noise}
    \vspace{5pt} 
  \end{minipage}
  \begin{minipage}{0.21\textwidth}
    \centering
    \includegraphics[width=\textwidth]{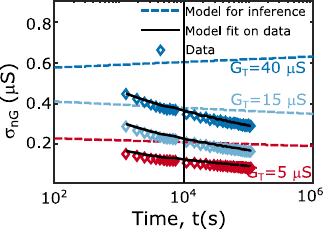}
    \subcaption{Read noise.}
    \label{fig:Read_Noise}
    \vspace{5pt} 
  \end{minipage}
  
  
  \caption{PCM device noise, taken from \cite{joshi2020accurate}.  All of these non-idealities are  state-dependent.}
  \label{fig:noises}
  \vspace{-15pt}
\end{figure}

To facilitate the implementation of PCM devices in synaptic networks, they can be configured as a differential pair allowing for the storage of both positive and negative weight. Such differential-pair cells arranged in a crossbar fashion enable in-memory computing of matrix-vector multiplication operation \cite{joshi2020accurate,papandreou2011programming}.  Fig.~\ref{fig:noises} depicts the noise obtained from measurements of over 10,000 devices in \cite{joshi2020accurate}.  PCM devices exhibit the following forms of non-idealities:

 



\begin{itemize}
    \item Programming noise $\sigma_p(G)$ -- also called write noise as it causes deviation from desired conductance upon writing to memory -- is attributed to the inherent stochasticity of crystallization process \cite{doi:10.1063/1.5042408}. These variations are approximately Gaussian distributed with a state-dependent standard deviation.
    \item The read noise $\sigma_r(G)$, due to fluctuations arising from $1/f$ noise \cite{doi:10.1063/1.5042408,LEGALLO202063}, is of smaller magnitude as compared to the programming noise.
    \item Conductance drift causes a reduction of PCM conductance over time. This is due to structural relaxation phenomenon affecting primarily the amorphous phase of the material \cite{doi:10.1063/1.5042408, gibbs1983activation}, with the drift rate also being a state-dependent parameter 
    \cite{joshi2020accurate}.
\end{itemize}

\section{Architecture and Design}
\label{sec:AnD}
In this section, we discuss the features of Bayes2IMC, our proposed architecture illustrated in Fig.~\ref{fig:First}. We first introduce the mathematical techniques that allow the utilization of device stochasticity as a source of noise distribution in Sec.~\ref{sec:arch_overview} and the corresponding hardware implementation in Sec.~\ref{subsec:lev_stoch}. We then describe the details of the core operations of the proposed Bayes2IMC architecture (Sec.~\ref{sec:single_core_operations}) and introduce the proposed output logit correction scheme (Sec.~\ref{sec:post-proc-logit}). Finally, Sec.~\ref{sec:Drift_Comp} discusses the drift compensation mechanism.
\subsection{Overview}\label{sec:arch_overview}

To generate a sample of the network weight in hardware, we need two key elements:
\begin{enumerate}
    \item Parameters that fully describe the weight distribution. For binary BNNs, this amounts to storing one parameter per weight, the probability parameter $p_w$ of the weight $w$ where, $p_w = \mathrm{Pr}(w=+1)$.
    \item A noise source that generates a standard distribution, e.g., the uniform distribution $\mathcal{U}(0,1)$ or the Gaussian distribution $\mathcal{N}(0,1)$. 
\end{enumerate}
The sample weight is then generated by comparing the parameter $p_w$ with a random variable $r\sim \mathcal{U}(0,1)$\cite{katti2024bayesian}, i.e.,
\begin{equation}
    w = 
     \begin{cases}
        +1 &\quad \text{if }r \leq p_w,\\
       -1 &\quad \text{if } r > p_w.\\
     \end{cases}
    \label{eq:unibinar}
\end{equation}
These sample weights make up the layers of the network mapped across cores, with each core performing the MVM operation
\begin{equation}
    y^{c}_{i} = \sum_{j=1}^M x^{c}_j w^{c}_{ji},
    \label{eq:mvmbinar}
\end{equation}
where  $\boldsymbol{x}^{c} = [x^{c}_1, ..., x^{c}_M]^\top$ is the core input, $w^{c}_{ji}$ are the binary weights indexed by $j$ and $i$ in the weight matrix and $y^{c}_i$ is the $i^{th}$ element of output vector $\boldsymbol{y^{c}}$.

The parameters $\boldsymbol{p}_{\boldsymbol{w}} = \{p_w\}_{w \in \boldsymbol{w}}$ required for sample generation are stored in WP. The noise, provided by the NP, is due to the inherent device-to-device and cycle-to-cycle stochasticity represented by the state-dependent programming noise shown in Fig.~\ref{fig:noises}. This Weight Plane-Noise Plane (WP-NP) crossbar array architecture introduced in \cite{10181438},  assigns $M$ rows for the weight probability parameters,  and $L$ rows in the NP to generate the required stochasticity. We reuse the hardware resource for sampling so that $L \ll M$, and empirically show that this does not affect network accuracy, leading to improved area efficiency. Noise cells from NP and weight cells from WP combine to generate $w$. We now elaborate   the details of this sampling utilizing device stochasticity.

\begin{table}
    \centering
\caption{Glossary of variables used in the text}
\label{tab:glossary}
    \begin{tabular}{ccc}\hline
         \textbf{Variable}&  \textbf{Definition}& \textbf{Values/Remarks}\\\hline\hline
         $w$&  Binary weight& $w\in \{+1,-1\}$\\
         $p_w$&  Probability parameter for $w$& $p_w \in [0,1]$\\
         $\lambda_w$&  Natural parameter for $w$& $\lambda_w \in( -\infty,\infty)$\\
         $z_w$&  Reparametrized $p_w$& $z_w \in( -\infty,\infty)$\\
 $\lambda_{clip}$& Clipping value for $\lambda_w$&$\pm 3.3$\\
 
         $z_{clip}$&  Clipping value for $z_w$& $\pm3$\\
         $f_{\sigma_{p}}$&  Programming noise model& as per Fig.~\ref{fig:Prog_Noise}\\
 $\kappa$& $z_w$ to $G_w$ scale&$8$\\
 $n_r$& No. of NP rows  read in parallel&$n_r \in \{1,2\}$\\
 $T_{WP}$& WP read pulse width&-\\
 $T_{NP}$& NP read pulse width&-\\\hline
    \end{tabular}
\end{table}
\subsection{Leveraging Device Stochasticity }\label{subsec:lev_stoch}

\noindent\emph{Reparametrization}: 
To generate network parameters as per \eqref{eq:unibinar}, we require a source of Uniform noise.  However, the naturally available stochasticity from the PCM-based NP in the array is Gaussian distributed. 

\begin{figure}
         \centering
         \fbox{\includegraphics[width=0.25\textwidth]{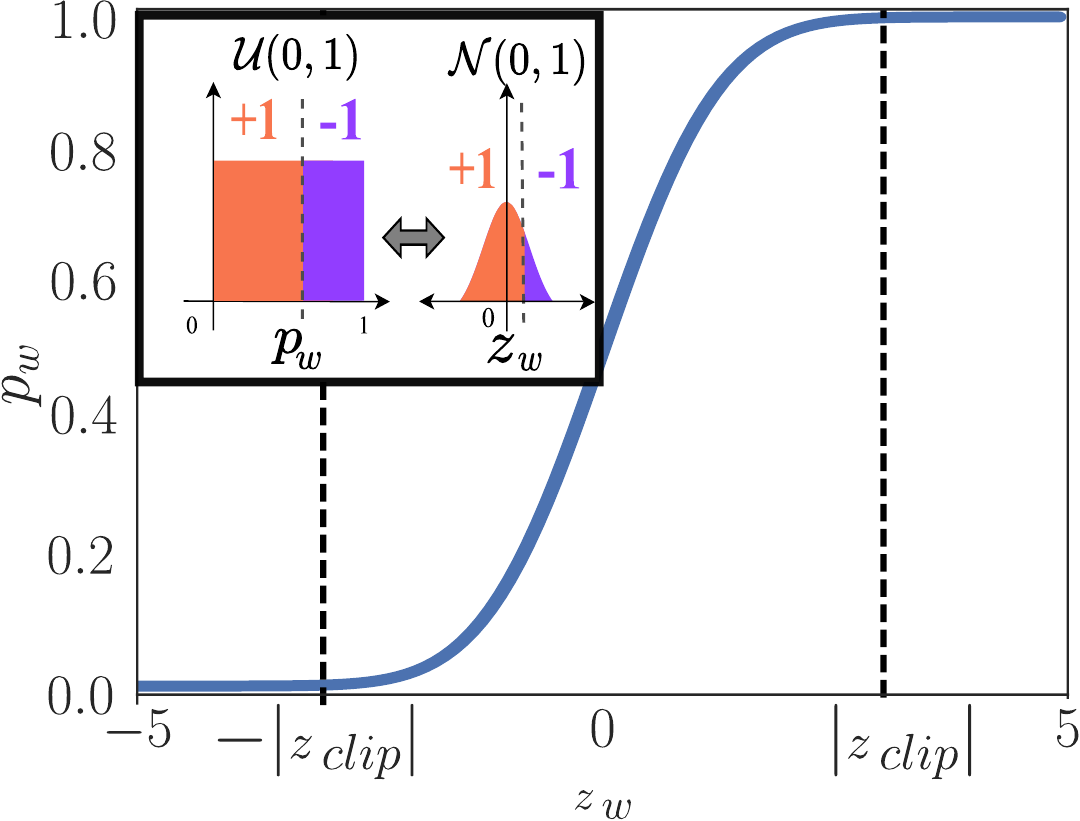}}
         
     \vspace{2pt}
     
         \fbox{\includegraphics[width=0.44\textwidth]{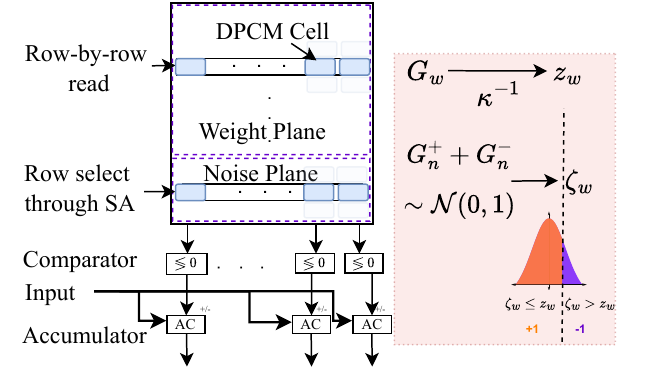}}
         \caption{\textbf{Top:} Probability parameter $p_w$ as a function of $z_w$. $z_w$ is clipped to $|z_{clip}|=3$. The error in $p_w$ due to these clippings in negligible. \textit{Inset:} Illustration of equivalence of Uniform sampling with $p_w$ and Gaussian sampling with $z_w$. \textbf{Bottom:} Detailed illustration of weight sampling as per \eqref{eq:normbinar}, and MVM operation as per \eqref{eq:core_equation}. }
  \label{fig:repar_and_core_ops}
    \vspace{-15pt}
\end{figure}

Here, we develop a principled method for sampling binary weights from Gaussian distribution using transform sampling \cite{press2007numerical}. Accordingly, we reparametrize or transform the probability parameter $p_w$ into another parameter, $z_w$ such that it can be  used with Gaussian noise for weight sampling.  The reparametrization is performed as
\begin{equation}
    z_w = \Phi^{-1}(p_w),
    \label{eq:phiinv}
\end{equation}
where $\Phi$ is the cumulative distribution function (CDF) of $\mathcal{N}(0,1)$ and $\Phi^{-1}$ is its inverse.
As illustrated in Fig.~\ref{fig:repar_and_core_ops}, the each weight $w$ is then independently sampled as
\begin{equation}
    w = 
     \begin{cases}
        +1 &\quad \text{if } \zeta \leq z_w,\\
       -1 &\quad \text{if } \zeta > z_w,\\
     \end{cases}
    \label{eq:normbinar}
\end{equation}
with $\zeta \sim \mathcal{N}(0,1)$.

We use the differential pair configuration of PCM devices, a standard method for representing positive and negative parameters, as described in  \cite{joshi2020accurate,9401384,khaddam2021hermes,aguirre2024hardware}. Each differential PCM (DPCM) cell, whether in WP or NP has two devices, programmed to conductances $G^+$ and $G^-$. Let the $\sigma_p(G)$ and $\sigma_r(G)$ denote programming and read noise for a given device with conductance $G$. We denote the conductance of a cell in weight plane storing parameter $z_w$ as $G_w$, and the conductance of the noise plane cell as $G_{n}$.

\noindent\emph{Noise source}: As opposed to the standard implementation of DPCM weights, here we programme the NP cells such that that the approximate equality $G_{n}^{+} \approx G_{n}^- \approx G_{n}$ holds. This ensures that the mean conductance read out from this cell almost vanishes, while the programming noise component from the devices adds up.  Programming noise has the dominant effect on the total noise of the noise plane cell as it is significantly larger than read noise as per Fig.~\ref{fig:noises}. Accordingly, the generated noise parameters are approximated as 
\begin{align}
    \mu_{n} &= G_{n}^{+} - G_{n}^{-} \approx 0,  \label{eq:mean1}\\
    \sigma_{n}^2 &= \sigma_{p}^2 (G_{n}^{+}) + \sigma_{p}^2 (G_{n}^{-})+\sigma_{r}^2 (G_{n}^{+}) + \sigma_{r}^2 (G_{n}^{-})\\
    & \quad \approx 2\sigma_{p}^2 (G_{n}).\label{eq:mean2}
 \end{align}


Our goal is to generate  $\zeta\sim \mathcal{N}(0,1)$ from each noise cell, as per \eqref{eq:normbinar}. Therefore, $G_{n}$ must be chosen such that the $\sigma_{n}$ as per  \eqref{eq:mean2} is equal to $1$, and can be identified from the programming noise model $f_{\sigma_p}$  shown in Fig.~\ref{fig:Prog_Noise}. 

\noindent\emph{Mapping parameters to WP}: We  now discuss the mapping strategy of a WP cell, with weight $w$ and the parameter  $z_w$.  We first scale up $z_w$ by a factor $\kappa$ to get $G_w$ which allows the  full use of the available conductance dynamic range of the device. IMC also reduces the impact of both programming and read noise in WP, ensuring that only the NP noise determines  the generation of the weight samples  \cite{10181438}. 

 The dynamic range available to the DPCM cell is [0,25$\,\mu$S], which means that the $|z_w|$ has to be clipped to some maximum value (denoted $|z_{clip}|$). For our experiments, we choose $|z_{clip} = 3|$ and map to networks after scaling it by $\kappa=8$, ensuring that we can map the entire required range of $0\leq p_w \leq 1$. 
 Empirically, we observed that $\lambda_w$, the natural parameter obtained post-training (related to $p_w$ as per \eqref{pwlamdaw}) can vary over several orders of magnitude. To ensure numerical stability, we clip $\lambda_w$ before transforming it to $p_w$ and subsequently to $z_w$. 

\noindent\emph{MVM operation}: Consider a weight matrix $\boldsymbol{W}^{c}$ with a corresponding parameter matrix $\boldsymbol{Z}^{c}$ with dimensions $N\times M$. The real-valued elements of  $\boldsymbol{Z}^{c}$ is mapped to the analog condutances of the devices in WP of the crossbar with $M$ rows and $N$ columns. Corresponding to each column, we provision  $L$  NP cells, which are read based on a stochastic arbitration scheme to generate the required noise source. Hence, the NP portion of the crossbar is organised with $L$ rows and $N$ columns. The stochastic arbitration scheme is employed to reduce the effect of correlation among the weights in a column. 

The MVM operation is accomplished in a row-by-row fashion.  A row of WP, denoted by index $j$ is read along with a stochastically selected row of NP, indexed by $j'$, to generate one row of $\boldsymbol{W}^{c}$, $\boldsymbol{w}^{c}_j \in \{-1,+1\}^{N\times 1}$. Simultaneously, the $j^{th}$ element of the input vector to the core $x^{c}_j$, is streamed into all column accumulators,  to be accumulated with the sign determined by each element of $\boldsymbol{w}^{c}_j$. This is repeated for each input element  $x^{c}_j$ by stepping through all the $M$ rows in the WP to complete the MVM operation given in \eqref{eq:mvmbinar} and obtain output vector $\boldsymbol{y^{c}}$. We denote $G_{w^{c}_{ji}}$ for the conductance corresponding to $w^{c}_{ji}$ and $G_{n^{c}_{j'i}}$ for the corresponding NP cell.
Mathematically, the MVM operation discussed in \eqref{eq:mvmbinar} is realised in hardware as 
\begin{equation}
    y^{c}_i = \sum_{j=1}^N x^{c}_j \times\mathrm{sgn}(\kappa^{-1}\big(G^+_{w^{c}_{ji}}-G^-_{w^{c}_{ji}}\big) + G_{n^{c}_{j'i}}^{+} - G_{n^{c}_{j'i}}^{-} ).
    \label{eq:core_equation}
\end{equation}
The stochastic arbitration is carried out by a dedicated control unit employing relatively lightweight pseudorandom generators such as linear feedback shift registers (LFSR) \cite{saluja1987linear}. The sample generation and MVM operation are illustrated in Fig.~\ref{fig:repar_and_core_ops}.

\subsection{Single Core Operations}
\label{sec:single_core_operations}
\begin{figure}[h]
    \centering
    \begin{minipage}[c]{0.48\textwidth}
        \includegraphics[width=\textwidth]{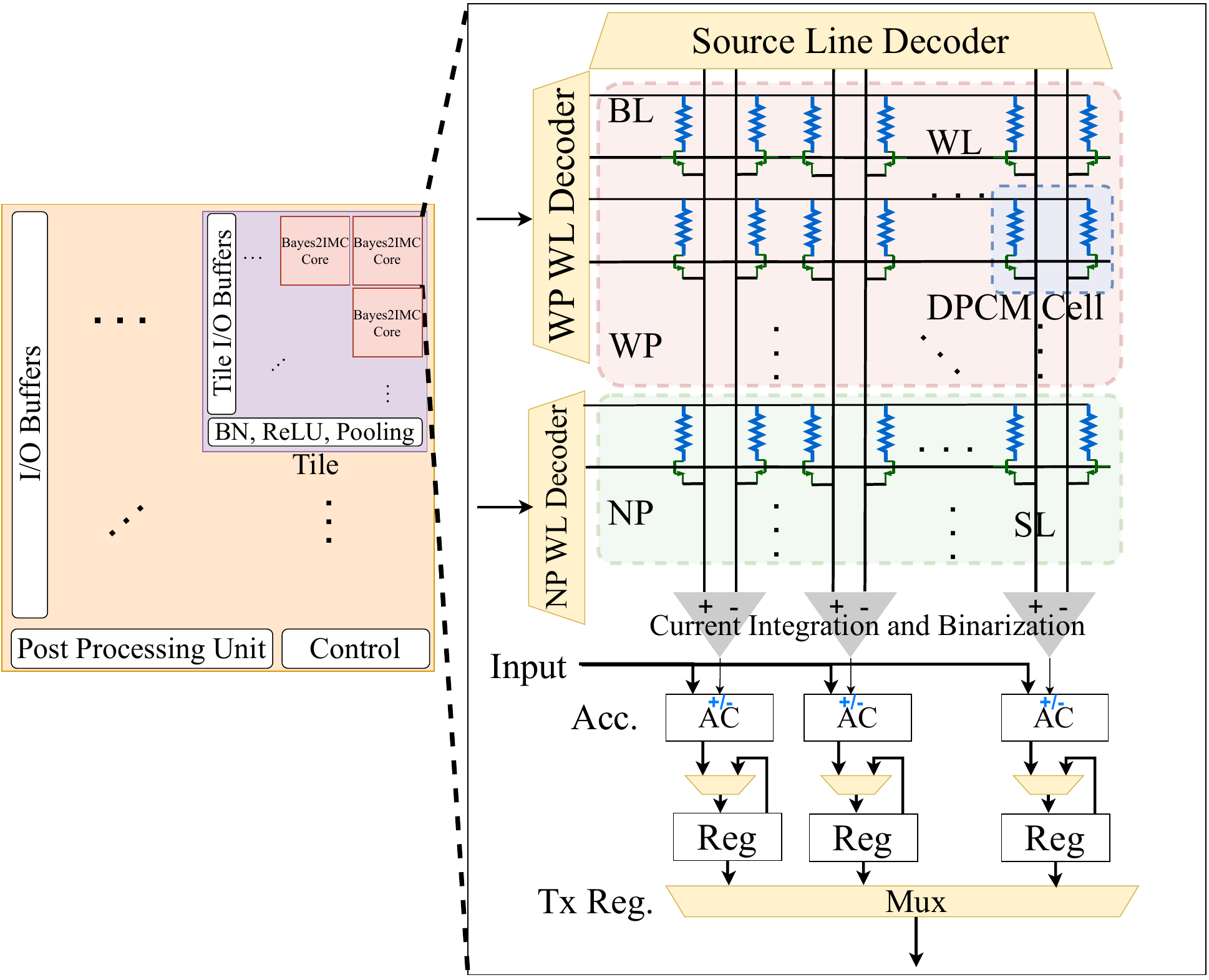}
      \end{minipage}  
      \hfill
        
    \caption{\textbf{Left:} A multi-tile core block diagram, with each tile composed of the Bayes2IMC cores along with peripherals such as BN unit, ReLU, pooling and buffers. Post-processing unit acts on the final layers for logit correction described in Sec.\ref{sec:post-proc-logit}. }\textbf{Right:} Detailed core design, expanding on Fig.~\ref{fig:First} and \ref{fig:repar_and_core_ops} illustrating all the elements in the crossbar and the periphery. 
    \label{fig:Complete_Core_Elements}
\end{figure}
We now discuss the details of the inference operation of a single DPCM crossbar core (Fig.~\ref{fig:Complete_Core_Elements}).  The size of the crossbar in the core is $144 \times 128$, with $128$ rows dedicated to WP and $16$ rows dedicated to NP. The NP row size of  $16$ was deemed to be sufficiently large to balance the need for reduced correlation amongst the parameters during sampling and area and power efficiency of the core \cite{10181438}. Correlation was further reduced by utilizing stochastic arbitration, as discussed in Sec.~\ref{subsec:lev_stoch}. 

Each of these  areas of the crossbar have a different mechanism for row-select, therefore they each have a dedicated decoder.
A DPCM cell consists of two PCM devices each with its own transistor selector arranged as a differential pair, in the standard 1T1R configuration.

\noindent\emph{Write operation}: PCM devices are written or programmed through an iterative programming scheme described in \cite{joshi2020accurate}. This involves a program-and-verify loop where programming pulses are modulated based on the conductance read iteratively until the device reaches close to the desired conductance value. A single ADC-DAC pair is enough for this scheme -- the output of the current-to-voltage converter shown in Fig.~\ref{fig:Complete_Core_Elements} can be routed to ADC input via a multiplexer (this multiplexer is different from the one shown in Fig.~\ref{fig:Complete_Core_Elements}), and the DAC access can be provided via the decoders. The source line (SL) multiplexing ensures that the other devices are not switched on when a device is being programmed. The write circuitry  can be power-gated during the inference operations, and will not contribute to the inference power. By provisioning only one ADC-DAC pair per core, the total area cost of incorporating the write circuitry can also be further reduced.

\begin{figure}

    \centering
    \includegraphics[width=0.5\textwidth]{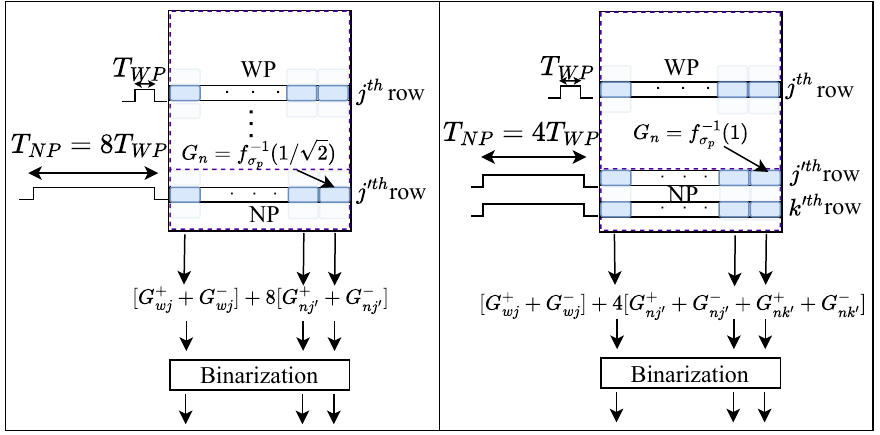}
    \caption{Two modes of read operation --  \textbf{Left:} Standard mode with $T_{NP}=8T_{WP}$ with $n_r=1$. \textbf{Right:} The high throughput scheme (right) that has $T_{NP}=4T_{WP}$ with $n_r=2$.}
  \label{fig:4vs8}
    \vspace{-15pt}
  \end{figure}

\noindent\emph{Input and read operation}: 
To execute  synaptic sampling, a WP and NP row are read together so that the resultant currents are summed up by Kirchoff's law. To this end, the bit lines are first precharged and held at the read voltage of the PCM device. Then the selected WL in WP is turned on along with the world line selected from NP using stochastic arbitration.  This enables the summation of the WP and NP currents on the SL, whereupon it is amplified to a rail-to-rail voltage signal. This signal is sent to the accumulator (AC) to determine the sign of $x_j$.
In order to implement the $\kappa$ scaling required for the WP and NP devices as in \eqref{eq:core_equation}, we vary the duration of the WL selector signal, as illustrated in Fig.~\ref{fig:4vs8}. The duration of the turn-on pulse applied to the NP row ($T_{NP}$) is scaled by a factor $\kappa$ compared to the duration of the pulse applied to the WP row ($T_{WP}$). While this has an impact on the achievable throughput, using amplitude-based encoding is challenging as it requires area and power-intensive DACs \cite{aguirre2024hardware} and also due to the non-linear relationship between read pulses and device conductivity \cite{7092504}.
 
However, to boost the achievable throughput, we  evaluate operational schemes that involve turning on several NP rows in parallel. If $n_r$ rows are turned on, the duration of the applied turn-on pulse in the NP can be reduced proportionately to $T_{NP}/n_r$, as long as the stored conductance values in the NP devices are also scaled appropriately as illustrated in Fig.~\ref{fig:4vs8}. We discuss the performance with $n_r=1$ and $2$ and find that $n_r > 2$ is not feasible as the device noise available from PCM is insufficient to achieve the desired amount of noise required for synaptic sampling. 

 Current integrator is then used to accumulate the current from WP and NP. A current integrator design adopted from \cite{cai2019fully} and \cite{aguirre2024hardware} is depicted in Supplementary Section C. The source line currents are  accumulated into integrating capacitors, one each for the two SLs corresponding to the differential pair. The output voltages of the capacitors thus developed are then binarized using a differential sense amplifier,  thereby implementing \eqref{eq:core_equation}. 

\noindent\emph{MAC}:  During the synaptic sampling operation, the 8-bit elements of input vector $x^c_j$ are streamed one-by-one to the corresponding 16-bit accumulators. The accumulator is responsible for summing up the individual contributions of input-weight dot product in $M$ accumulation cycles as described in \eqref{eq:core_equation}. This is accomplished by adding or subtracting $x_j$ from the partial MAC value based on its sign determined by $w_{ji}$. 
This way, we avoid power- and area-hungry multi-bit ADCs.
Upon accumulating the pre-neuron output for all the rows described in \eqref{eq:core_equation},  they are then sent to the transfer registers that hold the pre-neuron outputs as they are transferred out of the core via the bus to which all the column outputs are multiplexed. Transfer registers allow the pipelining of the whole operation -- while the previous pre-neuron outputs are waiting to be transferred out, the next set of MVM operations are simultaneously underway in the accumulator. 

\noindent\emph{Multi-core operation}: Since most typical layers in neural networks are larger than the size of a Bayes2IMC core ($144\times128$), a layer is usually mapped across multiple cores \cite{peng2019dnn+}. We refer to a collection of such cores as a tile. Partial sums of such cores are then carried out across all the cores. The partial sums are then routed into a tile-level neuron processing unit performing pre-activation accumulation, Batch Normalization (BN), ReLU, and max-pooling. 

\noindent\emph{Batch normalization and ReLU}: In conventional implementations, BN operations are folded into convolutional layers \cite{perez2021heterogeneous} as MVM operations are linear and BN only performs an affine transform on the products. In our case, however, as a non-linear operation (binarization) is required before MAC, it is necessary to perform a BN operation separately after the MVM operation. Having a BN unit per column was found to be costly because BN operations involve large digital multipliers.  
Besides, when a larger network is mapped into many cores, as is often the case, only a small number of cores perform BN operations, and the majority transfer the partially accumulated sum to the tile accumulator. Therefore we employ one BN multiplier and adder unit at each tile, significantly reducing area and power.
BN coefficients can be stored in a small memory array, either built with SRAM or with PCM devices with each device storing only one-bit \cite{9401384}. The latter is more area and power-efficient, and because it is stored in a binary fashion by programming the devices to $\{0, G_{max}\}$ which also ensures sufficient margin between the two levels. 

\noindent\emph{Post inference processsing}: The entire software-trained network can be thus deployed across various such cores of the Bayes2IMC architecture. Once the results of the operations from the last layer are available, they are streamed out of the core to perform logit corrections, followed by softmax to obtain probabilities in a dedicated unit off-tile. 
The entire operation is repeated $N_{MC}$ multiple times to get an ensemble of predictions (i.e ensembling-in-time).

\noindent\emph{Frequentist networks}: We note that the same core design can be easily utilized to perform frequentist inference. The weights $|w|$ can be scaled to $|G_{max}|$ before mapping. In such a case, we need not earmark parts of the crossbar for NP, and instead use all the cells to store weights $\boldsymbol{w} \in \{-1,+1\}^{|\boldsymbol{w}|}$, and subsequently binarize and accumulate as described above.

\subsection{Post-Processing Logit Correction}
\label{sec:post-proc-logit}
 Due to device variations, $z_w$ will be inaccurately programmed onto the WP. Committee machines \cite{joksas2020committee} have been proposed to mitigate the effects of these imperfections through ensemble averaging. However,  since there is only one copy of each parameter in the ensembling-in-time paradigm adopted in this work, the programmed parameter remains unchanged across all ensembles (apart from the small effects of read noise), leaving no scope for committee-machine style parameter error smoothing. 
This could cause unpredictable corruption of logit distribution at the output, resulting in large variations in accuracy across multiple deployments. Therefore, we need a hardware-software co-optimized calibration operation that can be applied after mapping the network to hardware to obtain stable accuracy. We describe a logit correction scheme that is applied only on the output of the last layer during post-processing towards this.

We focus on an $n$-class classification problem and define the logits $l(k|\boldsymbol{x},\boldsymbol{w}) \in \mathbb{R}$ for an input $x$ and for each label $k \in \mathscr{C}$, such that the predicted label probability is given by the softmax of the logit vector $\boldsymbol{l}(\boldsymbol{x},\boldsymbol{w}) = [l(0|\boldsymbol{x},\boldsymbol{w}), ..., l(n-1|\boldsymbol{x},\boldsymbol{w})]^\top$. 
For any label $k \in \mathscr{C}$, the logits obtained from the available data $(\boldsymbol{x}, y) \sim \mathcal{D}$ under model $\boldsymbol{w}$ during the pre-deployment phase are empirically observed to exhibit a bi-modal distribution consisting of two unimodal distributions $\hat{p}(L_k | y=k)$ and $\hat{p}(L_k | y \neq k)$, where $L_k = l(k|\boldsymbol{x},\boldsymbol{w})$ is the random variable representing the empirical logit distribution on $\boldsymbol{x} \sim \mathcal{D}$. 
Both $\hat{p}(L_k | y=k)$ and $\hat{p}(L_k | y \neq k)$ are modeled as Gaussian distributions with parameters $\theta_{k,1} =(\mu_{k, 1}, \sigma_{k, 1})$ and $\theta_{k,0} =(\mu_{k, 0}, \sigma_{k, 0})$ respectively.
This choice is grounded in previous works, which demonstrate that deep Bayesian neural networks converge to Gaussian processes under certain conditions \cite{pmlr-v48-gal16, williams1996computing, agrawal2020wide}, and further reinforced by empirical observations of logit distributions on our datasets.

\begin{figure}
\centering
\includegraphics[width=0.43\textwidth]{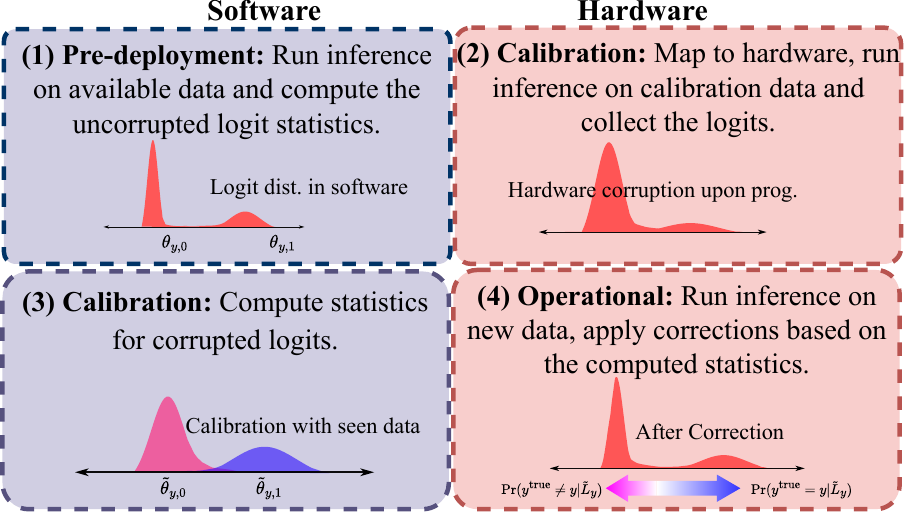}
\caption{ A hardware-software co-optimized flow for logit correction.}
\label{fig:hwswcopt}
\end{figure}

As described before, the non-idealities of the PCM devices lead to distorted logit distributions. The hardware distortions lead to reduction of resolvability of the predictions $p(y |\boldsymbol{w}, \mathcal{D})$. This results in blurring of the modes, causing degradation of separability. 
The distorted distribution, with its random variable  $\Tilde{L}_k$, also exhibits a bimodal structure with Gaussian distributions $\hat{p}(\Tilde{L}_k | y=k)$ and $\hat{p}(\Tilde{L}_k | y \neq k)$. These distributions are parametrized by $\Tilde{\theta}_{k,1} = (\Tilde{\mu}_{k, 1}, \Tilde{\sigma}_{k, 1})$ and $\Tilde{\theta}_{k,0}=(\Tilde{\mu}_{k, 0}, \Tilde{\sigma}_{k, 0})$, respectively.
Since the degree of corruption cannot be predicted in advance, we developed a post-facto algorithm that corrects these distortions after the network has been mapped to hardware, illustrated in Fig.~\ref{fig:hwswcopt}.

During the calibration phase, we use a subset $\mathcal{D}^{\mathrm{cal}} = \{x^{\mathrm{cal}}_i, y^{\mathrm{cal}}_i\}_{i=1}^{N^{\mathrm{cal}}}$ of the available data, called calibration data, to run inference on the programmed network and estimate the Gaussian parameters $\Tilde{\theta}_{k,1}$ and $\Tilde{\theta}_{k,0}$. 
We assume a one-to-one correspondence between $L_y$ and $\Tilde{L}_y$ as an affine perturbation, such that both variables differ by scale and offset factors depending on parameters $\theta_{k,0}$, $\theta_{k,1}$, $\Tilde{\theta}_{k,0}$, and $\Tilde{\theta}_{k,1}$. Let us define the affine correction estimate for each mode $k$ as  
\begin{equation}
 \mathbb{E}[L_k | \Tilde{L}_k, y] 
    = \frac{(\Tilde{L}_k - \Tilde{\mu}_{k,c})}{\Tilde{\sigma}_{k,c}}\sigma_{k,c}+\mu_{k,c},
    \label{eq:harddiscr}
\end{equation}
where $c = \delta_{\{y = k\}}$. 

The correction is then applied by weighting the affine correction estimate by the estimated probability of $\{y = k\}$ (depicted in Fig.~\ref{fig:hwswcopt} panel (4))  using Bayesian update as
\begin{align}
\hat{l}_k 
    &= \mathbb{E}[L_k | \Tilde{L}_k] \\
    &= \mathrm{Pr}(y = k | \Tilde{L}_k) \mathbb{E}[L_k | \Tilde{L}_k, y = k] \nonumber\\
    &\quad + \mathrm{Pr}(y \neq k | \Tilde{L}_k) \mathbb{E}[L_k | \Tilde{L}_k, y \neq k].
     \label{eq:1d}
\end{align}
Here, the posteriors $\mathrm{Pr}(y = k | \Tilde{L}_k)$ and $\mathrm{Pr}(y \neq k | \Tilde{L}_k)$ are calculated using the Bayes' theorem, with likelihood obtained from the value of Gaussian distribution function parametrized by $\Tilde{\theta}_{k,1}$ and $\Tilde{\theta}_{k,0}$ respectively at $\Tilde{L}_k$.

For a balanced dataset, prior probabilities $\mathrm{Pr}(y = k)$ and $\mathrm{Pr}(y \neq k)$  are equal to $\frac{1}{n}$ and $\frac{n-1}{n}$, respectively. Further details are provided in Supplementary Section B.
Accordingly, the posterior probabilities $\mathrm{Pr}(y = k | \Tilde{L}_k)$ and $\mathrm{Pr}(y \neq k | \Tilde{L}_k)$ determine the strength of correction application (scale and shift) towards either side, decreasing the fuzziness between the two modes. The process is repeated for each of the $N_{MC}$ predictors in the ensemble. As these operations are performed only on the output layer, they can be handled by a dedicated computational block off-tile, amortizing the overall cost.

\subsection{Drift compensation} \label{sec:Drift_Comp}
In this section, we discuss techniques to compensate the effects of drift. As discussed earlier, due to structural relaxation, PCM devices exhibit a state-dependent drifting of their conductance values. The observed conductance drift has an exponential dependence on time, with the drift exponent being state-dependent \cite{joshi2020accurate}. Mathematically, the device conductance at time $t$  is given as 
\begin{equation}
    G(t)=G(T_0)\bigg(\frac{t}{T_0}\bigg)^{-\nu}, \label{eq:drift}
\end{equation}
where $G(T_0)$ is the conductance at time $T_0$ after programming,  and $\nu$ is the state dependent drift exponent. Here, $T_0 = 20$\,s is the time after programming at which the noise characteristics  are defined as per \cite{joshi2020accurate}. 

Two compensation mechanisms for drift compensation were suggested in \cite{joshi2020accurate} --  Global Drift Compensation (GDC) and Adaptive Batch Normalization Scheme (AdaBS). These methods require periodic calibration applied to each layer of the network with a subset of the known data. In the former, the MVM output of a column is measured at various time instances and divided with the outputs obtained immediately after the instantiation of the array, to gauge the extent of compensation required.  The compensation factor thus calculated is used to scale the measured MVM products to reverse the effect due to drift. The latter uses running batch statistics instead of MVM outputs to calculate the compensations per layer. Both these methods measure the MVM output to estimate the extent of drift. This is necessary as individual weights drift at state-dependent rates, making a uniform, one-size-fits-all compensation impossible.
\begin{figure}[h!]
  \centering
    \includegraphics[width=0.45\textwidth]{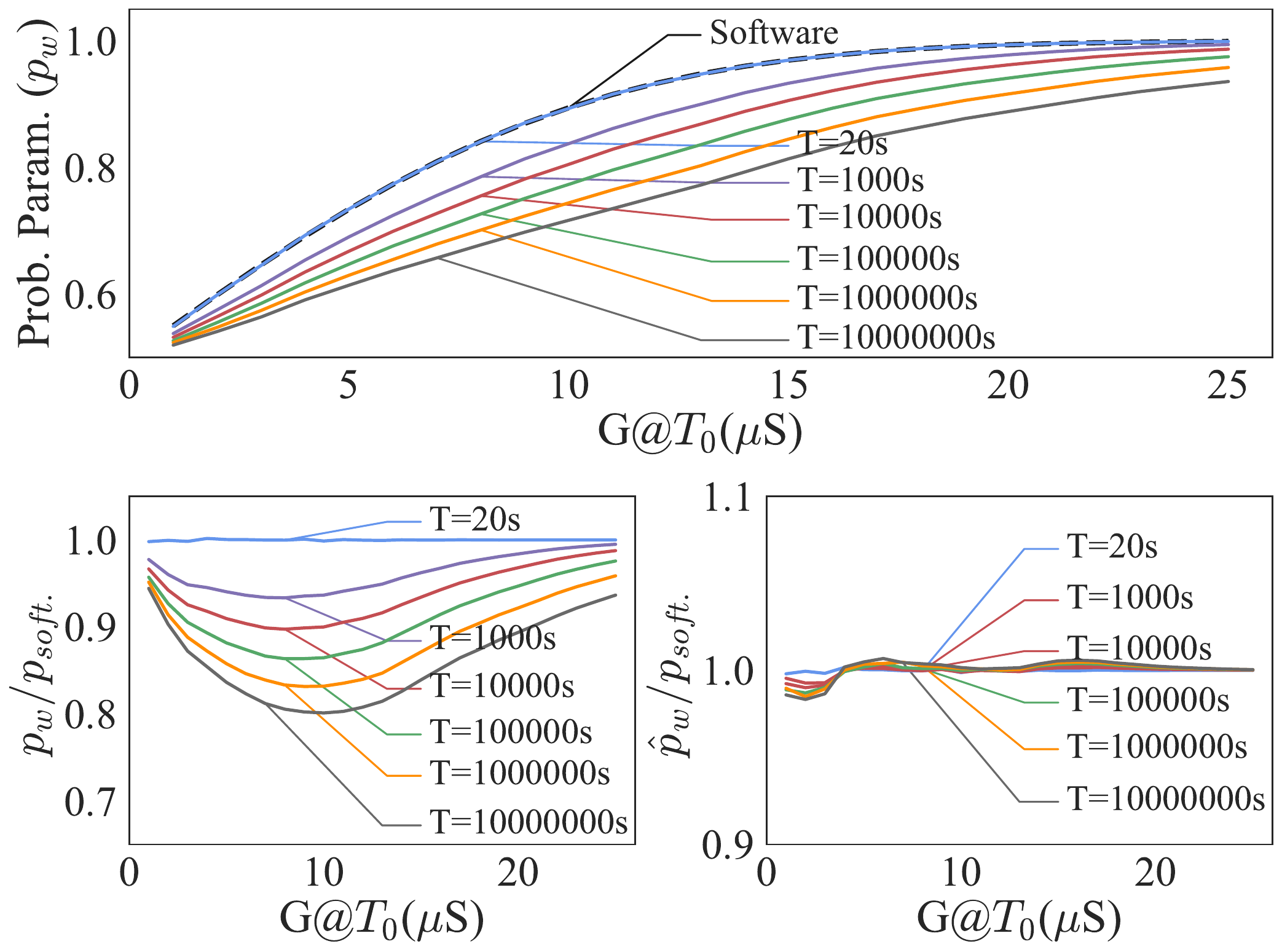}
    \caption{\textbf{Top}: $p_w$, the probability parameter for a weight $w$ vs  WP conductance G at $T_0 = 20$\,s. 
    \textbf{Bottom, Left}: Ratio of $p_w$ and $p_{soft.}$, the ideal, software calculated value of $p_w$. \textbf{Bottom, Right}: Ratio of $p_w$ by $p_{soft.}$ after applying correction $\alpha_t$.}
  \label{fig:drift_probs}
\end{figure}

We  describe how the reparametrization techniques introduced in this paper enable a straightforward and universal drift compensation method applicable to all the layers of any BBNN that does not have these limitations. Our approach, Bayes2IMC drift compensation (BBDC),  involves using a single drift correction coefficient   $\alpha_t = \big(\frac{t}{T_0}\big)^{\nu_c}$  to scale the duration of the read pulses applied to measure the device  conductances to reverse drift effects. 



To select the appropriate value of $\nu_c$, we examine how drift affects $p_w$. Since the reparametrized probability parameter $z_w$  is directly proportional to the conductance state, it is significantly influenced by drift. However, the impact of drift on $p_w$ is much less pronounced, as illustrated in Fig.~\ref{fig:drift_probs}.  We observe that conductance values near the extremes of our range ($0$ and $25 \mu$S) are  less affected by drift. Therefore, we select $\nu_c = 0.06$, corresponding to the drift exponent at approximately $G = 8\mu$S, and apply this correction across the entire network. This choice effectively minimizes errors in $p_w$ within the mid-conductance ranges while causing minimal changes at the extremes. As a result, the reparametrization of $p_w$ into $z_w$ enables a straightforward drift compensation across the entire conductance range. The effect of compensating factor $\alpha_t$ is illustrated in Fig.~\ref{fig:drift_probs}, which shows the relationship between $p_{soft.}$, the software calculated ideal probability parameters, $p_w$, and $\hat{p_w}$, the probability parameter after compensation. 

 The effect of drift on noise sourced from NP was empirically observed to be smaller than the magnitude of WP conductance. Therefore, we perform drift compensation by appropriately adjusting $T_{NP}$ to adjust the relative durations of the read pulses applied to WP and NP. Since $T_{NP}$ can only be integral multiples of $T_{WP}$, our scaling is quantized. 
 We show in section \ref{subsec:drift_performance} that this quantization has no impact on accuracy performance, and we can effectively correct for the accuracy degradation caused by drift. We also note that $T_{NP}$ reduces with time. As the throughput of the Bayes2IMC core depends on $T_{NP}$ since each parameter read operation needs that long to complete, we achieve a throughput speedup with time. This, however, comes at the cost of overall inference power consumption, as discussed in section \ref{sec:Results}.

 Thus, by calculating a single multiplying factor, $\alpha_t$, at time $t$ based on the drift behavior of PCM devices and then adjusting the read pulses, we can effectively mitigate the effects of drift. Our approach is robust and does not require calibration with known data to correct drift degradation, making it more versatile. However, the logit correction described in sec. \ref{sec:post-proc-logit} may still require such calibration. When combined with drift compensation, logit correction can further enhance accuracy and ECE stability. 

\section{Experiments and Results}
\label{sec:Results}
In this section, we evaluate the performance of the proposed Bayes2IMC architecture. We first train a VGGBinaryConnect network on CIFAR-10, following the standard practice in the literature for benchmarking   Bayesian neural networks \cite{bonnet2023bringing,lu2022algorithm,10.1109/DAC18074.2021.9586137}. VGGBinaryConnect is a modified version of a 9-layer frequentist VGG network with binary synapses described in \cite{Alizadeh2018AnES} and  \cite{10.5555/2969442.2969588}. IMCs BBNN counterpart was trained using the BayesBiNN optimizer described in \cite{Meng_Bachmann_Khan_2020}. Adopting the hardware-aware training methodology introduced in \cite{joshi2020accurate}, the network parameters are trained by injecting state-dependent  Gaussian noise representative of PCM programming stochasticity, thus avoiding the need for expensive on-chip training. For benchmarking, the frequentist version of the same network was trained using a Straight-Through Estimator (STE) with ADAM optimizer, as described in \cite{Meng_Bachmann_Khan_2020}.
The BBNN network was then transferred to a custom PCM hardware simulator implementing the noise characteristics described in Fig.~\ref{fig:noises} as well as appropriate input and accumulator quantization.  Inference was then performed with $N_{MC}=10$. The logit correction method discussed in sec. \ref{sec:post-proc-logit} were calibrated with $2000$ validation set images. All the results were reported from 6 different runs for each scenario, with each run comprising of (in case of a hardware implementation) an initial device programming step followed by inference. Accuracy, ECE, and uncertainty quantification results were obtained under various scenarios. Drift performance was evaluated for up to $T=10^7$\,s. Finally, hardware projections were calculated for the Bayes2IMC architecture and compared with equivalent SRAM architecture and existing implementations in the literature.

\subsection{Accuracy and ECE Performance}\label{sec:ACC_ECE_performance}

\begin{figure}
  \centering
    \includegraphics[width=0.5\textwidth]{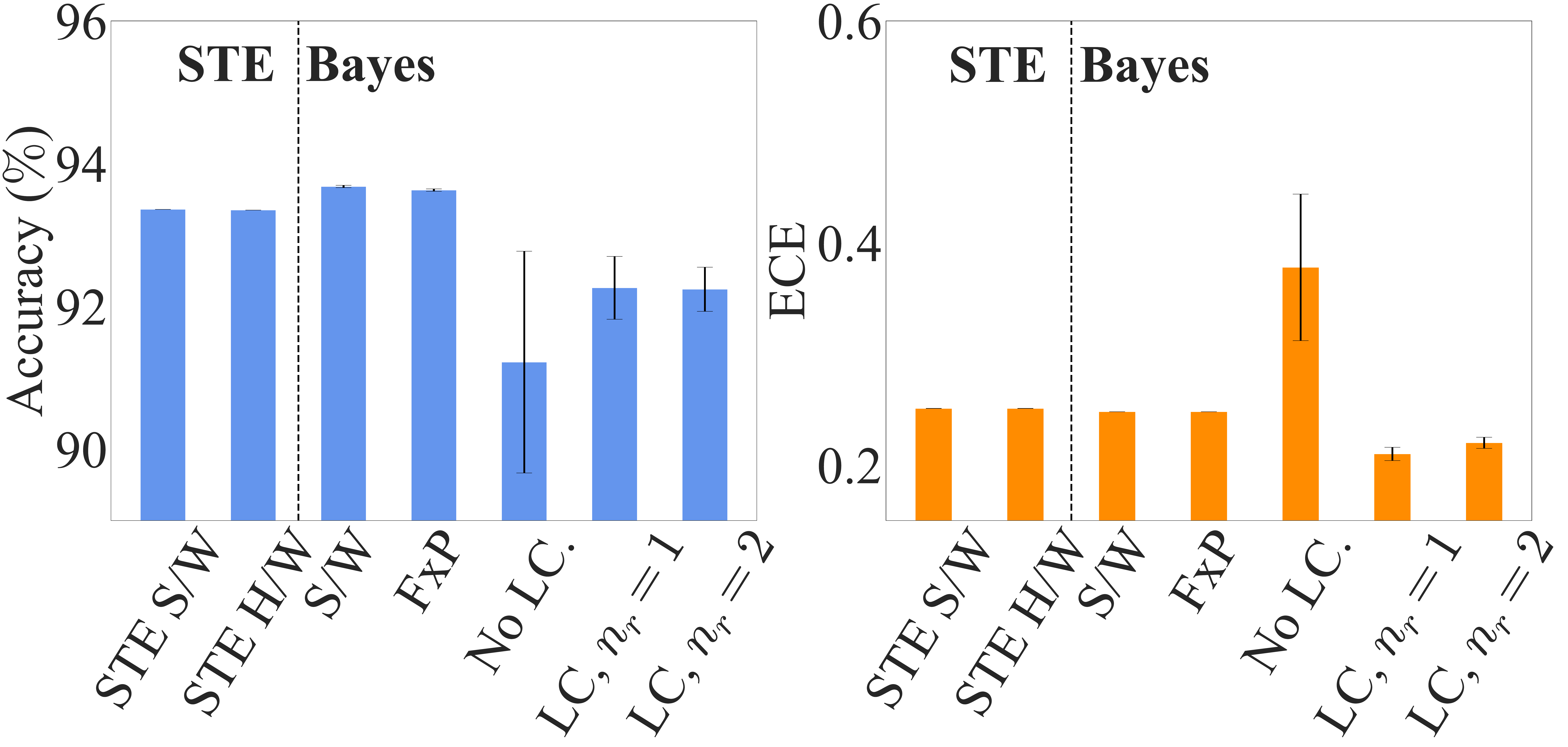}
    \caption{Accuracy and ECE on CIFAR-10. The first two bars on the left of the dashed line correspond to the frequentist-STE based network  while the rest correspond to the BBNN. LC refers to the logit correction. The error bars indicate $\pm 1$ standard deviation (SD).}
  \label{fig:accece}
\end{figure}

Accuracy and ECE evaluated on hardware (with logits corrected and uncorrected) were benchmarked against FP32 software  BBNN (i.e. probability parameters are represented in FP32), a frequentist STE-based network, and a fixed point (FxP) implementation (Fig.~\ref{fig:accece}). The fixed point network was parametrized with $5$-bit parameters for a fair comparison with a PCM implementation, as the effective PCM bit capacity is reported to be $4$ \cite{wong2010phase} making DPCM capacity $5$-bits. The frequentist network was also transferred to PCM simulator for hardware simulation. 

  The software classification accuracy of BBNN ($93.68\%$\ for FP32 and $93.63\%$ for fixed-point) is slightly better compared to the STE-based frequentist network ($93.36\%)$.  This matches the accuracy reported in literature \cite{Meng_Bachmann_Khan_2020}.  The estimated ECE values for both are found to be comparable. 

There is no drop in accuracy or ECE for the frequentist STE  network post-transfer to hardware. This is expected because devices are mapped to only two conductance levels. In contrast, the uncorrected Bayes2IMC accuracy post-mapping are poorer ($91.22\%$) with   high cross-deployment variations ($\sim\pm1.5\%$). Upon applying corrections to the logits, the accuracy variations were brought down to $0.4\%$. The correction boosts accuracy while decreasing ECE as well, with accuracy reaching within $1.5\%$ of the ideal software   accuracy ($92.26\%$ vs $93.68\%$) and ECE slightly improving in comparison to the FP32 software implementation   ($0.21$ vs $0.25$).

We also compare the performance of the two read schemes introduced  in section \ref{sec:single_core_operations}; for both $n_r=2$ and $n_r=1$ we obtain similar accuracy and ECE. This validates the multiple NP row read method introduced to increase throughput.

\subsection{Uncertainty Quantification}

\begin{figure}[htbp]
  \centering
    \centering
    \includegraphics[width=0.24\textwidth]{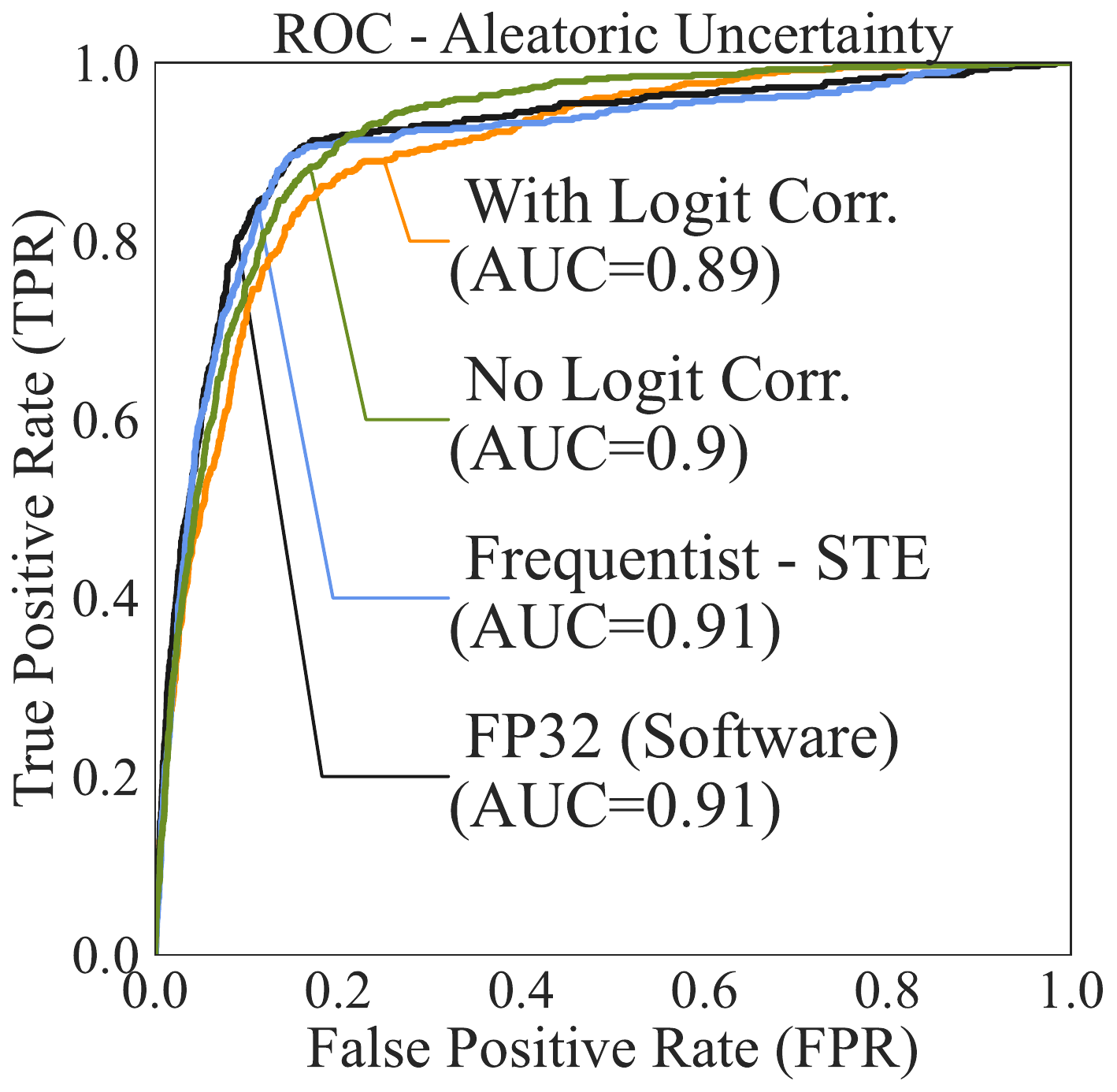} 
    \centering
    \includegraphics[width=0.24\textwidth]{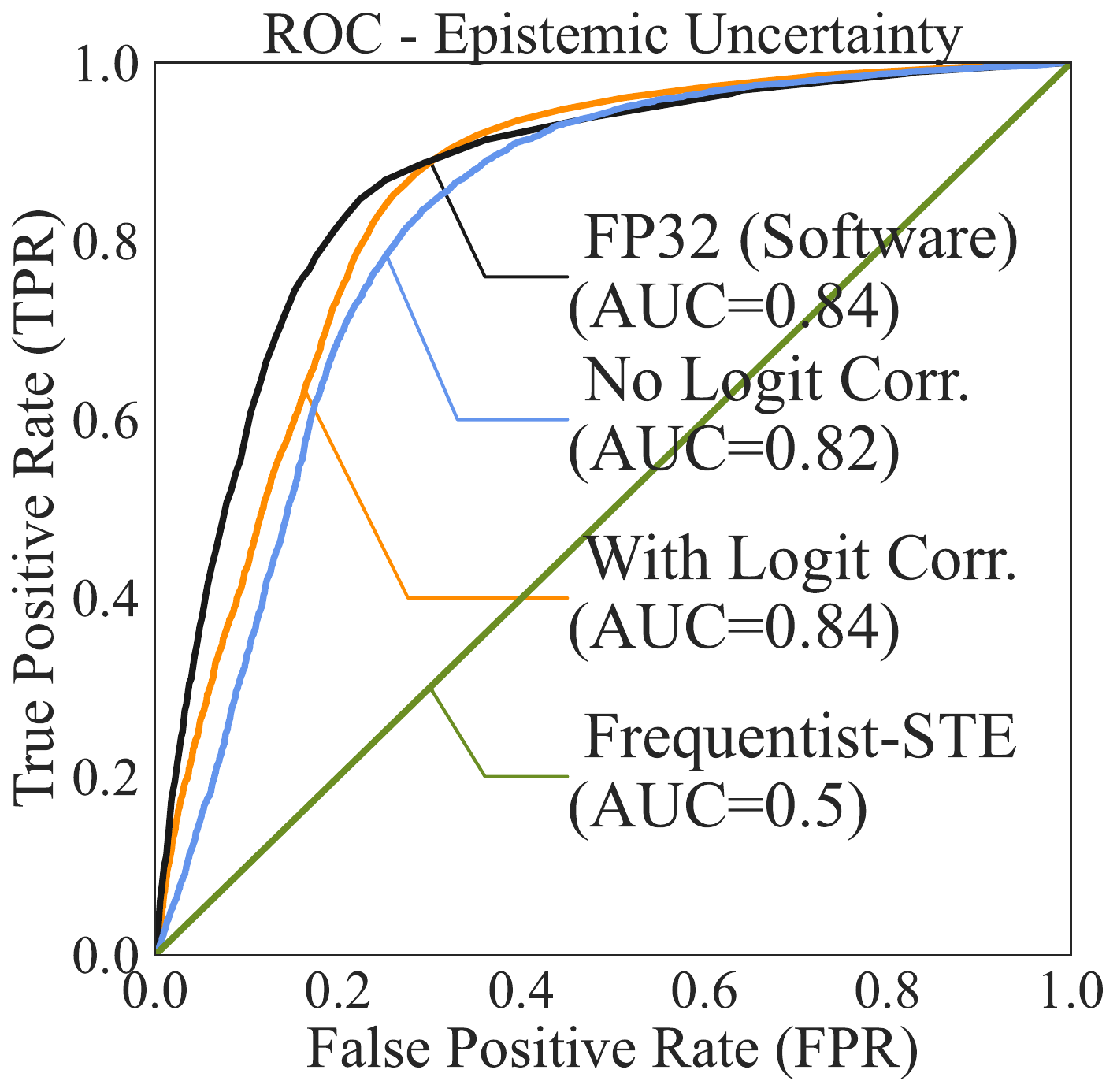} 

\caption{ROC of $U_a$ (Left) and $U_e$ (Right),   for FP32 (software), frequentist, uncorrected and corrected logit scenarios (hardware).}
  \label{fig:ROC}
  \vspace{-0.5cm}
  \end{figure}
  
Uncertainty is quantified by entropy measures as described in \eqref{eq:up}, \eqref{eq:ua} and \eqref{eq:ue}. For each input data, the aleatoric ($U_a$) and epistemic uncertainties ($U_e$) are computed. Aleatoric uncertainty ($U_a$) helps determine whether the prediction is correct or incorrect, while  epistemic uncertainty ($U_e$) identifies whether the input is IND or OOD, based on predefined thresholds. To evaluate the ability of the network to discriminate across all thresholds, we plot receiver operating characteristic (ROC) curves. For aleatoric uncertainty, the ROC curve measures the ability to differentiate between correct and incorrect predictions. For epistemic uncertainty, the ROC curve evaluates the ability to distinguish between IND and OOD data, with the CIFAR100 dataset used for OOD detection \cite{bonnet2023bringing}. The area under the curve (AUC) of each ROC is used as the metric to evaluate the uncertainty quantification performance.

We compared the performance of Bayes2IMC with uncorrected as well as corrected logits against software networks (Fig.~\ref{fig:ROC}). The AUC for both epistemic and aleatoric uncertainties for Bayes2IMC, with and without logit correction are found to be closely matching the FP32 software network AUC. The aleatoric uncertainty of the Bayes2IMC matches that of the frequentist-STE network, while the epistemic uncertainty AUC for the frequentist counterpart is $0.5$. This is because $U_e=0$ for all inputs for a frequentist network, indicating its inability to discriminate between OOD and IND data.




\subsection{Drift Mitigation }
\label{subsec:drift_performance}
\begin{figure}
  \centering
    \includegraphics[width=0.45\textwidth]{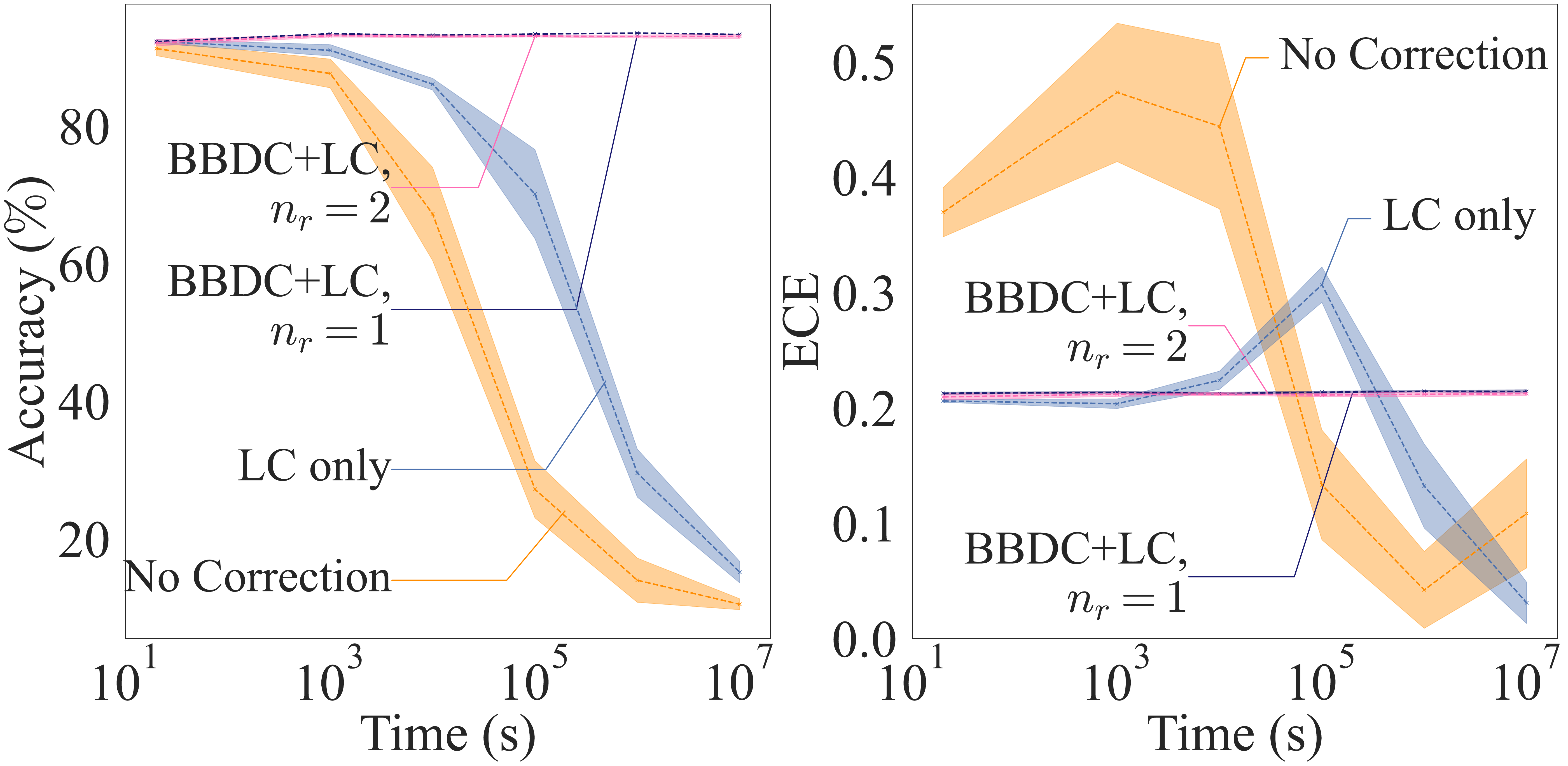}
    \caption{ Effect of drift on accuracy and ECE performance. The band around lines represent $\pm1$ SD.}
  \label{fig:drift_acc}
  \vspace{-0.2cm}
\end{figure}

\begin{figure}
  \centering
    \includegraphics[width=0.45\textwidth]{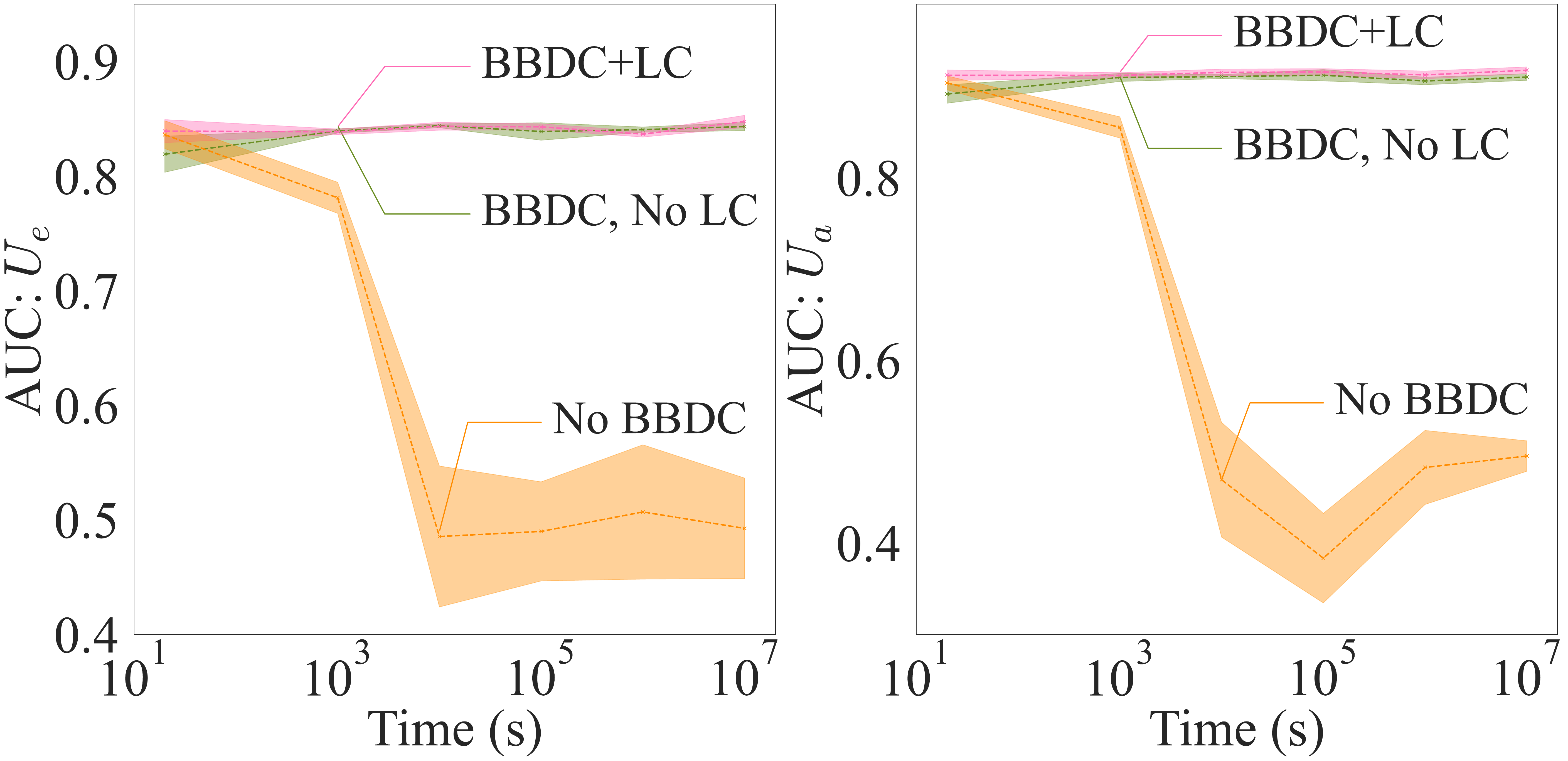}
    \caption{Epistemic and aleatoric uncertainty  due to effect of conductance drift under following scenarios: no drift compensation, drift compensation with logits uncorrected, and drift compensation with logit correction.} 
  \label{fig:drift_uncer}
  \vspace{-0.6cm}
\end{figure}
The effect of drift on accuracy and ECE was studied up to $T=10^7$\,s, as shown in Fig.~\ref{fig:drift_acc}. We observe that accuracy as well as ECE degrades significantly due to conductance drift. With no correction, the net accuracy of the network comes down nearly to $10\%$  in $10^6$\,s, which is equivalent to random guessing. In around $10^5$\,s, i.e., slightly longer than a day, the accuracy reduces to $20\%$. Logit correction slightly improve the performance -- with accuracy post correction at $10^5$\,s being around $70\%$.   
IMC is worth noting that it takes around $10^6$\,s for the accuracy of the output of Bayes2IMC to reduce to $10\%$ without any corrections. 
The reparametrization of $p_w$ to $z_w$ before mapping, rather than mapping $p_w$ directly, ensured significant robustness to drift, as seen in Fig.~\ref{fig:drift_probs}.

The sharp drop of ECE from $10^5$\,s onwards is accompanied by a severe decline in accuracy around the same time in both uncorrected and logit-corrected scenarios. This does not imply ECE performance improvement, because ECE only quantifies the divergence between accuracy and prediction confidence. Therefore, it is important to look at ECE in the context of the accuracy performance and not as a standalone metric.

Upon applying BBDC  technique introduced in this paper, we achieve a complete recovery of accuracy performance through $10^7$\,s. We also compare the uncertainty estimation performance of various correction methods in conjunction with BBDC.  This is accomplished by plotting AUC of ROCs for epistemic and aleatoric uncertainty across all time instances, as shown in Fig.~\ref{fig:drift_uncer}.  Without BBDC, the AUC for both $U_e$ and $U_a$ suffers significant performance degradation, collapsing to near $0.5$. With BBDC, the aleatoric and epistemic uncertainty performance completely recovers, and the results from both corrected and uncorrected logits closely match.

\subsection{Hardware Estimation}
\begin{table}
    \centering
    
\caption{SRAM vs PCM core}
\label{tab:hwperf}
    \begin{tabular}{>{\centering\arraybackslash}p{0.34\linewidth}>{\centering\arraybackslash}p{0.2\linewidth}cc} \hline
          &\textbf{Units}&   \textbf{PCM}&\textbf{SRAM}\\ \hline\hline
         Clock Rate &MHz& 
     100&208\\
 Ops/clk pulse ($8/4/2$)& -& 16/32/64&128\\ 
 Read power per weight&$\mu$W& 48.5&1993\\ 
 Total read power &mW& 6.2&256\\ 
 Total digital block power ($8/4/2$)&mW& 1.46/2.92/5.84&26.6\\ 
 \textbf{Total power}($8/4/2$)&mW& 7.6/9.1/12.0&282\\ 
 Total crossbar area&mm$^2$& 0.015&0.14\\ 
 Total sensing block area&mm$^2$& 0.02&0.04
\\ 
 Total digital block area&mm$^2$& 0.18&0.2\\ 
 Total BN memory area&mm$^2$& 0.002&0.007\\ 
 \textbf{Total area}&mm$^2$& 0.22&0.40\\ 
 \textbf{Power efficiency} ($8/4/2$)&GOPS/W& 208/350/531&94.6\\ 
 \textbf{Total efficiency} ($8/4/2$)&GOPS/W/mm$^2$& 941/1581/2397&250\\ 
 \hline   
    \end{tabular}
    \vspace{-0.3cm}
\end{table}

\begin{figure}
\vspace{-0.5cm}
     \centering
         \centering

         \includegraphics[width=0.45\textwidth]{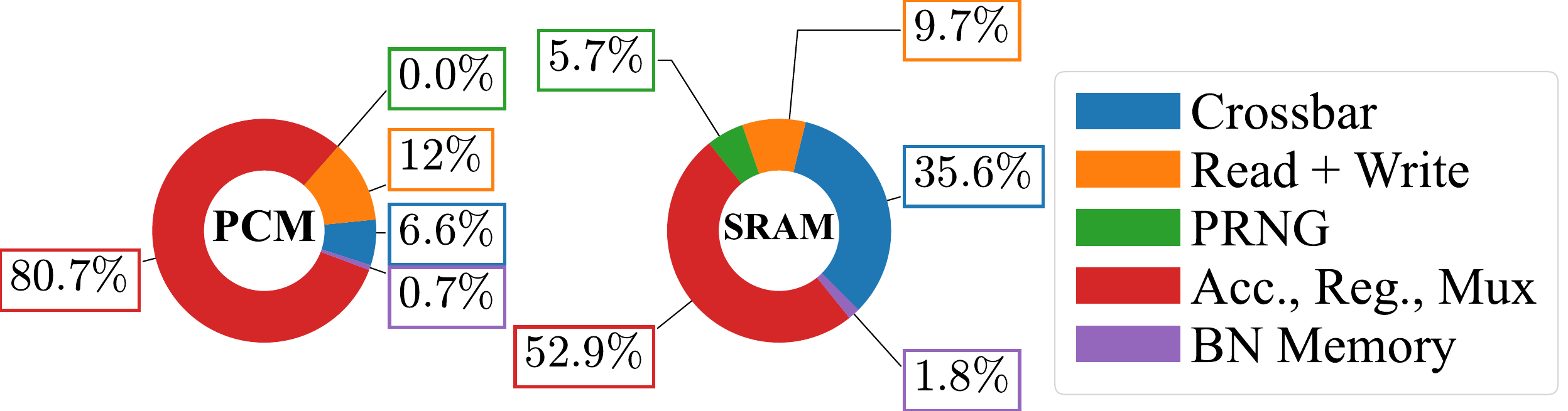}
     \caption{Area breakdown of a PCM and an SRAM core.}
  \label{fig:areabreakdown}

\end{figure}

We now estimate the hardware performance of the Bayes2IMC architecture in terms of throughput, area, and power of a single Bayes2IMC PCM core and compare it with an equivalent SRAM core with $5$-bit parameters.

Our simulations are based on large-array PCM hardware results reported in the literature by IBM Research where the memristive device was integrated with industry-standard 90 nm  CMOS technology. Hence, we consider 90 nm node for our hardware projections incorporating the device noise characteristics based on PCM cells fabricated in this node \cite{joshi2020accurate}. The cell read power and area for PCM devices used here are based on \cite{9116226}. SRAM read power is assumed to follow the trends reported in   \cite{pedram2016dark, datta2022ace}, and scaled to equivalent numbers for the 90 nm node using scaling laws described in \cite{stillmaker2017scaling}. Cell area of a single PCM was taken to be $29$\,F$^2$ per device as per \cite{9116226}, whereas that of a 6T SRAM-based cell was taken to be $120$\,F$^2$ \cite{peng2019dnn+}. We consider a $128 \times 128$ parameter storage per crossbar for both PCM and SRAM core. Additionally, $16$ rows are apportioned for NP, taking the total number of PCM core rows  to $144$. 
While each row read in SRAM requires only a single clock cycle, the   PCM architecture requires $8$ clock pulses for $n_r=1$ and $4$  clock pulses for $n_r=2$, as per Fig.~\ref{fig:4vs8}. For the $n_r=2$, the read pulse width can shrink to $2$ clock pulses at $10^7$\,s because of drift compensation.

The digital peripheral circuits, including accumulators, registers and multiplexers were synthesized in Cadence. The operating frequency for Bayes2IMC is assumed to be $100\,$MHz, based on the read pulse time required for the PCM device read as per \cite{9116226}. Since the total read time is determined by $T_{NP}$, the digital peripheral circuit is clocked at the same rate to synchronize operations. Thus, the speed-up of the core comes at the cost of read power. For an SRAM core, the frequency was calculated based on the timing report obtained from Cadence synthesis.

Fig.~\ref{fig:areabreakdown} shows the area breakdown of an PCM Bayes2IMC and the SRAM core respectively. In a Bayes2IMC PCM core, the digital peripheral circuitry—consisting of the accumulator, register, multiplexer block, and decoders—accounts for the majority of the overall area, making up $81\%$ of the footprint. In contrast, for SRAM, the crossbar area is comparable to the size of the peripheral area. Because of the dominance of peripherals, the total area of PCM core is only $2 \times$ smaller than that of SRAM even though the PCM crossbar is almost $10 \times $ smaller than its SRAM counterpart.

\begin{table*}[htpb]
    \centering
    
\caption{Comparison with other works in literature}
\label{tab:litcomp}
    \begin{tabular}{>{\centering\arraybackslash}p{0.2\linewidth}>{\centering\arraybackslash}p{0.1\linewidth}cccccc} \hline
          &\textbf{Units}&   \textbf{\cite{cai2018vibnn}}&\textbf{\cite{10.1109/DAC18074.2021.9586137}} & \textbf{\cite{yang2020all}}&\textbf{\cite{Jia2020EfficientCR}} & \textbf{\cite{lu2022algorithm}}&\textbf{This work}\\ \hline\hline
         Network&-& 
     MLP&ResNet-18& MLP& MLP& VGG-16&VGGBinaryConnect\\ 
 Dataset&-& MNIST&CIFAR-10& MNIST& MNIST& CIFAR-10&CIFAR-10\\ 
 Parameter distribution&-& Gaussian&Gaussian& Gaussian& Gaussian& Gaussian&Binary\\ 
 Weight precision&bits& 8&8& 8& 8& 8&1\\ 
 Accuracy on CIFAR-10&\%& -&92.9& -& -& 88&92.2\\ 
 Implementation&-& FPGA&FPGA& DW-MTJ& CMOS ASIC& SOT-MRAM&PCM\\ 
 Node&nm& 28&20& 28& 45& 22&90\\ 
 Throughput&GOPS& 59.6&1590& -& 1.86& 92&6.4\\ 
 Area &mm$^2$& -&-& -& 6.63& -&0.22\\ 
 Power&W& 6.11&45& & 0.5& 0.0089&0.012\\ 
 Power efficiency&GOPS/W& 9.75&35.33& 2516& 3.72& 10387&531\\
 Total efficiency& GOPS/W/mm$^2$& -& -& -& 0.5611&- &2568\\
 Power efficiency (90nm)& GOPS/W& 0.42& 0.91& 175.82& 0.86& 447.3&\textbf{531}\\
 Total efficiency (90nm)& GOPS/W/mm$^2$& -& -& -& 0.046& -&\textbf{2568}\\ 
 \hline

 \end{tabular}
 \vspace{-0.5cm}
\end{table*}
The digital peripheral circuitry in latter also consists of the PRNG unit, which was designed with maximal reuse architecture discussed in \cite{katti2024bayesian}, where all $4$-bytes on a $32$ bit LFSR were utilized for random number generation.  The PRNG unit makes up almost $10\%$ of total peripheral area and $6\%$ of the total core area. Additionally, we highlight the area required to store $128$ pairs of BN coefficients, one per column. Table \ref{tab:hwperf} also compares the power consumed by the SRAM and the Bayes2IMC core. Owing to the increase in clock rates due to the decrease in $T_{NP}$, we have reported the power for three modes, indicated as $8/4/2$ corresponding to $n_r=1$,  $n_r=2$, and for $T_{NP}/T_{WP} = 2$ at $t=10^7$ for $n_r=2$ respectively.

We evaluate efficiency for the three modes of operation (with read pulses $8/4/2$ indicated in Table \ref{tab:hwperf}) as discussed earlier. We find that the PCM core architecture has power efficiency (in Giga-operations per second per Watt or GOPS/W) of $2.2\times$ for $T_{NP}/T_{WP}=8$, $3.7 \times$ for $T_{NP}/T_{WP}=4$ and $5.6\times$ for $T_{NP}/T_{WP}=2$ mode, compared to an SRAM core. Overall efficiency (GOPS/W/mm$^2$) gains are respectively $3.8\times$, $6.3\times$, and $9.6\times$.

We also compare our work with other works on Bayesian neural network architectures reported in literature. These include implementations on CMOS (FPGA and ASIC), as well as on NVM devices such as DW-MTJ and SOT-MRAM utilizing nanoscale stochasticity. We compute the power efficiency in terms of  GOPS/W and total efficiency in terms of GOPS/W/mm$^2$. For fair comparison, we translate these numbers to 90 nm node equivalent using scaling laws described in \cite{stillmaker2017scaling}. We find that our design achieves one of the best accuracies on CIFAR-10 despite having single-bit weights. IMC also outperforms all CMOS architectures as well as the DW-MTJ (iso-node) implementation in terms of area and power efficiency. We also have 1.2$\times$ better  best-case ($T_{NP}/T_{WP}=2$) and comparable nominal case (0.78$\times$ @ $T_{NP}/T_{WP}=4$)  iso-node power efficiency compared to the recent state-of-the-art \cite{lu2022algorithm}. Note that the approach reported in\cite{lu2022algorithm} achieves high efficiency by performing ensembling only on selected layers, as well as by utilizing selector-less, high-speed,  SOT-MRAM cells.
\vspace{-0.35cm}
\section{Conclusion}
\label{sec:Conclusion}
This work introduced Bayes2IMC, a binary Bayesian neural network implemented with in-memory computing on standard PCM crossbar hardware. By leveraging the conductance variations of PCM devices —typically considered a drawback— for ensembling, we reduced the core peripheral area by nearly $10\%$ compared to an SRAM based core that utilizes a digital psuedo-random number generator. Our approach includes a reparametrization algorithm and an innovative scheme that eliminates the need for ADCs, significantly enhancing both power and area efficiency.

Through a hardware-software co-optimization flow, output logits were corrected to enhance accuracy and the ECE. Additionally, the reparametrization technique supports a straightforward drift compensation method that fully mitigates drift-induced accuracy degradation. We also successfully demonstrated uncertainty quantification and decomposition into aleatoric and epistemic uncertainties using Bayes2IMC.

Our hardware projections, informed by literature and CAD projections, indicate that while Bayes2IMC operates $2-8 \times$ slower than an SRAM-based architecture, it achieves up to $10 \times$ higher efficiency in terms of GOPS/W/mm$^2$. We also projected that Bayes2IMC is up to $20\%$ more power-efficient than the state-of-the-art.  Our analysis reveals that the digital and analog peripheral circuitery contributes significantly more to area and power consumption than the crossbar or memory elements themselves. Future efforts to improve efficiency must thus focus on optimizing peripheral circuits. The techniques, algorithms, and co-optimization framework discussed in this paper are demonstrated with PCM devices but are versatile enough to be applied to other noisy NVM device-based compute architectures for implementing binary Bayesian networks. 
\vspace{-0.6cm}
\section*{Acknowledgment}
 BR's research was supported in part by the EPSRC Open Fellowship EP/X011356/1 and by the EPSRC grant EP/X011852/1. OS's research was supported by an Open Fellowship of the EPSRC with reference EP/W024101/1, by the EPSRC project EP/X011852/1, and by the European Union’s Horizon Europe Project CENTRIC under Grant 101096379.
\vspace{-0.1cm}
\begin{IEEEbiography}
 [{\includegraphics[width=1.1in,height=1.25in,clip,]{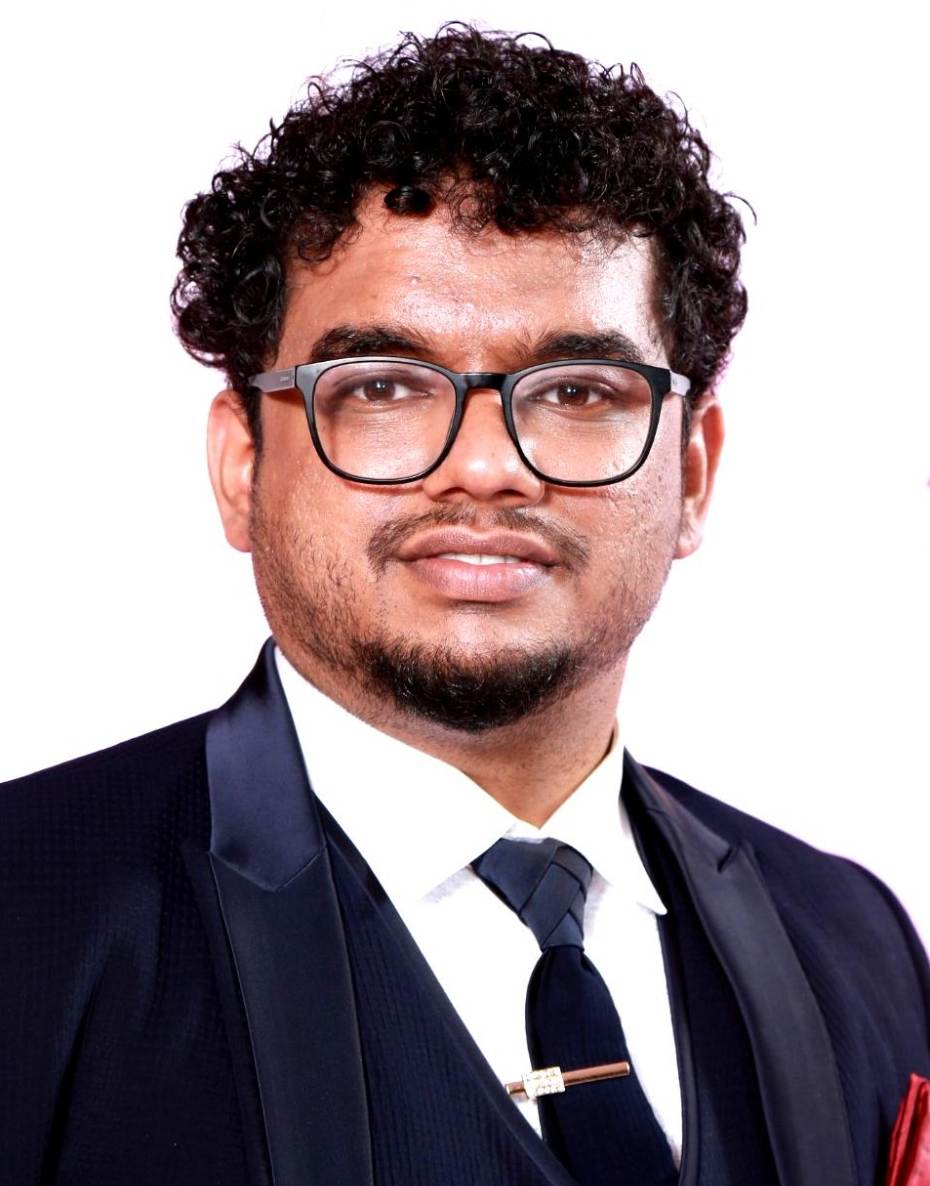}}]
{Prabodh Katti} (Graduate Student Member, IEEE) is a post-graduate research scholar in the Department of Engineering at King's College London. His interests include neuromorphic computing, Bayesian learning, and AI hardware. He obtained his M.Tech in Electronic Systems Engineering from the Indian Institute of Science (IISc), Bangalore, and B.Tech in Avionics from Indian Institute of Space Science and Technology (IIST), Thiruvananthapuram. He has worked in Indian Space Research Organization (ISRO), Ahmedabad, on payloads and instruments on Indian earth observation and planetary missions.
\end{IEEEbiography}

\begin{IEEEbiography}
[{\includegraphics[width=1.1in,height=1.25in,clip,keepaspectratio]{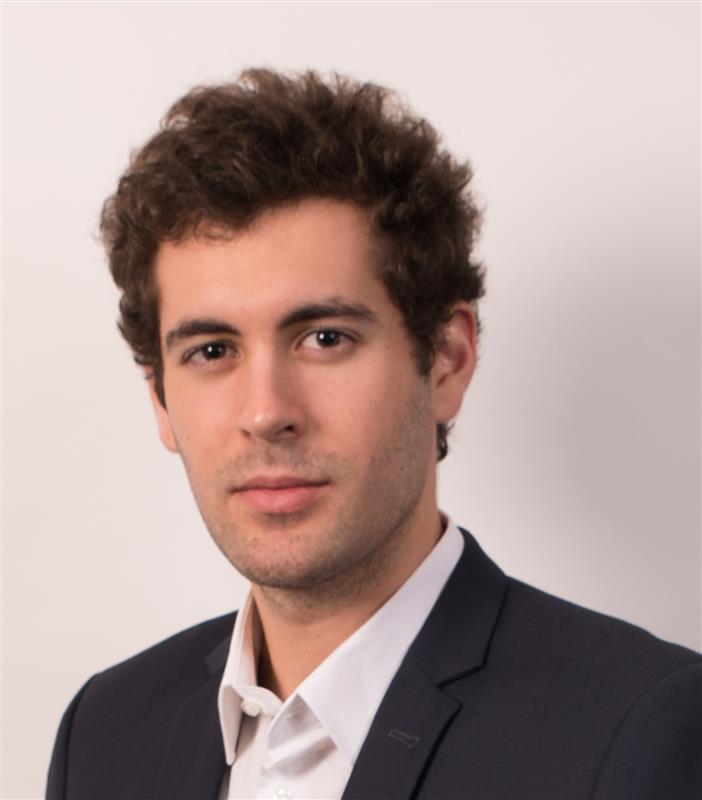}}]
{Clement Ruah} (Graduate Student Member, IEEE) received the first master’s degree in engineering from French “Grande École” CentraleSupélec and the second master’s degree (Hons.) in biomedical engineering (neuroscience stream) from Imperial College London in 2018. He is currently pursuing a Ph.D. degree in machine learning with King’s College London, U.K. His research interests include wireless ray-tracing, model-based reinforcement learning and Bayesian learning.

\end{IEEEbiography}
\vspace{-30pt} 

 \begin{IEEEbiography}    [{\includegraphics[width=1.1in,height=1.25in,clip,keepaspectratio]{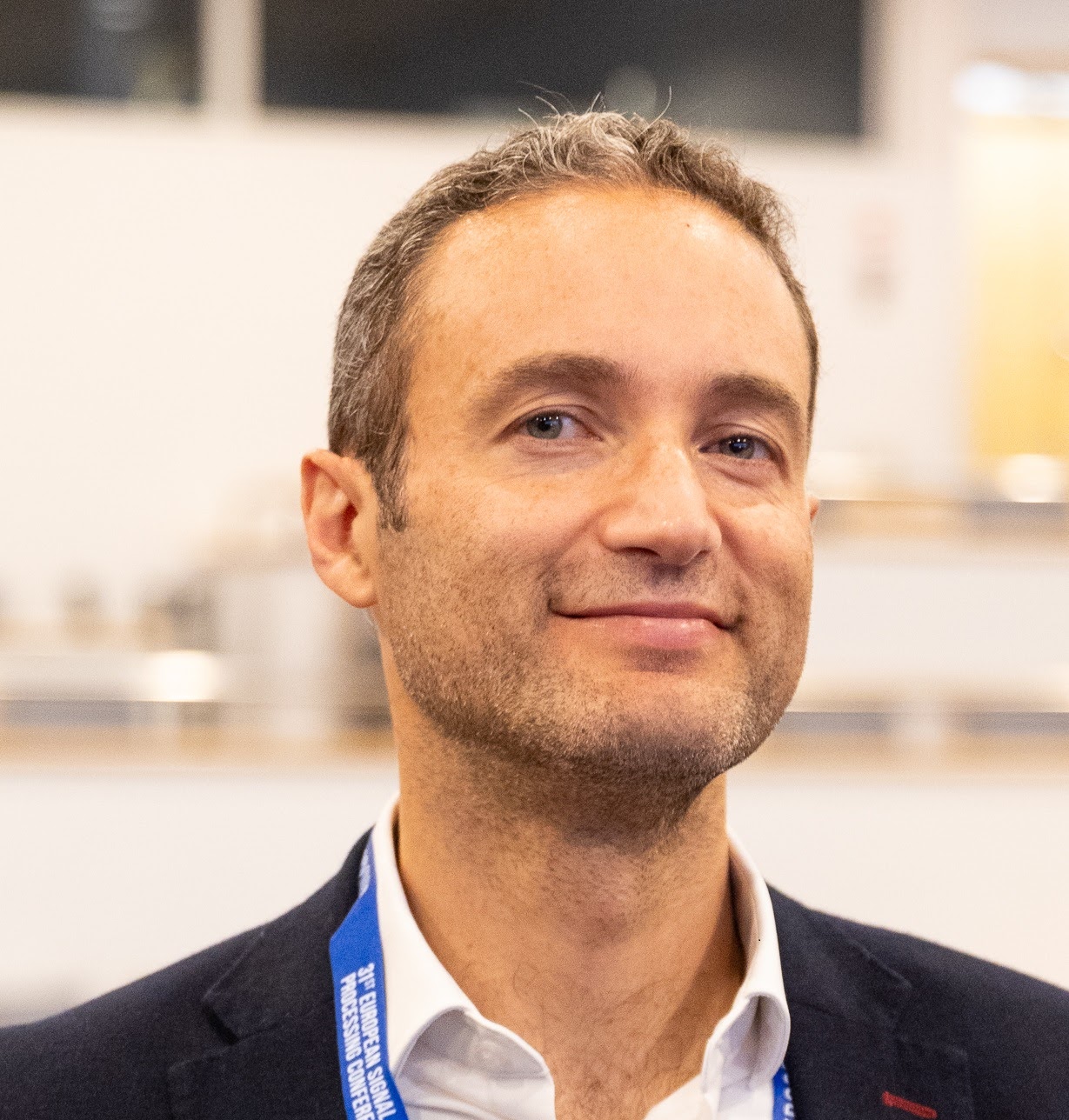}}]
 {Osvaldo Simeone} (Fellow, IEEE) is a Professor of Information Engineering, and he co-directs the Centre for Intelligent Information Processing Systems within the Department of Engineering of King's College London. He is also a visiting Professor with the Connectivity Section within the Department of Electronic Systems at Aalborg University. From 2006 to 2017, he was a faculty member of the Electrical and Computer Engineering (ECE) Department at New Jersey Institute of Technology (NJIT). Among other recognitions, Prof.  Simeone is a co-recipient of the 2022 IEEE Communications Society Outstanding Paper Award, the 2021 IEEE Vehicular Technology Society Jack Neubauer Memorial Award, the 2019 IEEE Communication Society Best Tutorial Paper Award, the 2018 IEEE Signal Processing Best Paper Award, and the 2015 IEEE Communication Society Best Tutorial Paper Award. He was awarded an Open Fellowship by the EPSRC in 2022 and a Consolidator grant by the European Research Council (ERC) in 2016. He was a Distinguished Lecturer of the IEEE Communications Society in 2021 and 2022, and he was a Distinguished Lecturer of the IEEE Information Theory Society in 2017 and 2018.   Prof. Simeone is the author of the textbook ``Machine Learning for Engineers"  published by Cambridge University Press, four monographs, two edited books, and more than 200 research journal and magazine papers. He is a Fellow of the IET, EPSRC, and IEEE.  
\end{IEEEbiography}
\vspace{-25pt}

 \begin{IEEEbiography}    [{\includegraphics[width=1.1in,height=1.25in,clip]{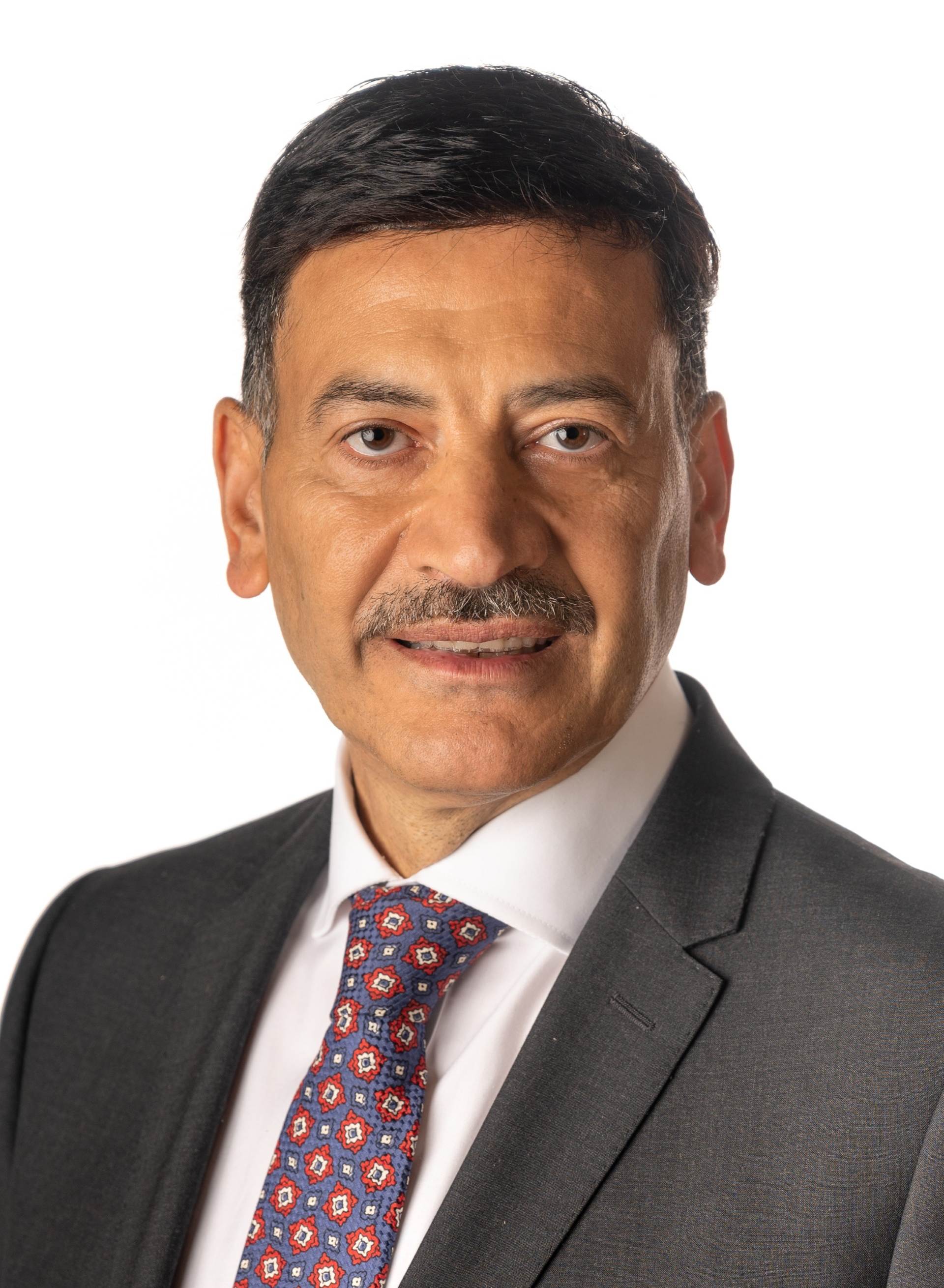}}]
{Sir Bashir M. Al-Hashimi} (Fellow, IEEE) is recognised for sustained, pioneering contributions to advanced semiconductor chips test, energy-efficient embedded systems and energy harvesting computing.  His research has led to substantive innovations, technology transfer and impact worldwide in academia and industry through related enabling hardware and software technologies applied in mobile electronic systems and devices. He is a highly cited researcher and has published nearly 400 technical papers, with eight best paper awards at international conferences and he has authored, co-authored and edited eight  books. He has successfully supervised 51 PhD theses to date and has secured over £25m in external research funding from UKRI and industry. He is the Vice President (Research and Innovation) and Arm Professor of Computer Engineering at King's College London. He was appointed CBE by HM the late Queen in 2018 for services to engineering and industry and knighted in HM King Charles III’s New Year Honours 2025 for his contribution to engineering and education. He was elected Fellow of the UK Royal Academy of Engineering (2013), to the Fellowship of The Royal Society (2023) and membership of the European Academy of Sciences and Arts (2023). He has also received several international awards for his research, notably including the IET Faraday Medal (2020).   
\end{IEEEbiography}
\vspace{-15pt}
 \begin{IEEEbiography}    [{\includegraphics[width=1.1in,height=1.25in,clip]{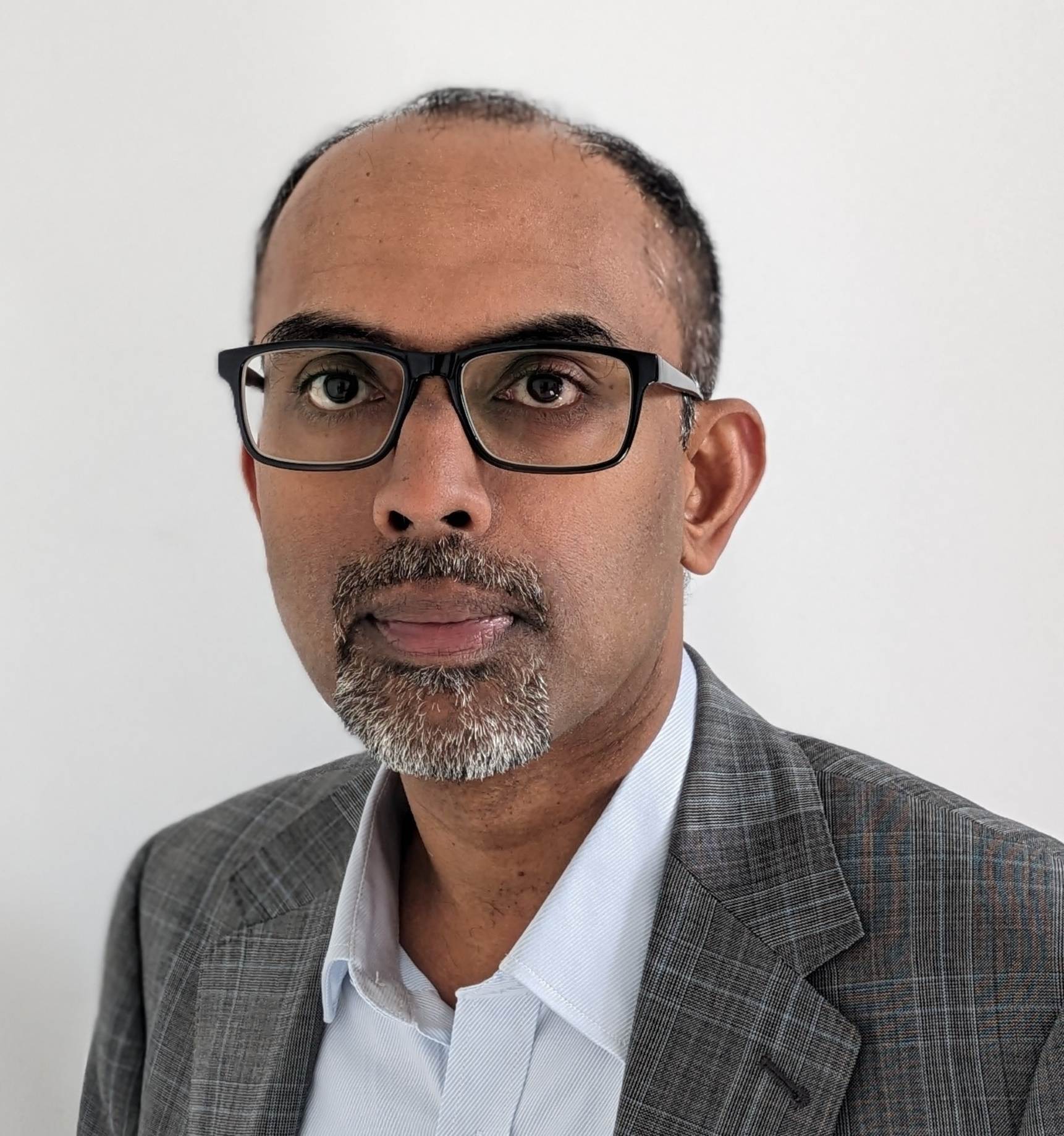}}]
{Bipin Rajendran}(Senior Member, IEEE) is a Professor of Intelligent Computing Systems and an EPSRC Fellow at King’s College London (KCL). He received his B.Tech from IIT Kharagpur in 2000 and M.S. and Ph.D. in Electrical Engineering from Stanford University in 2003 and 2006, respectively. From 2006 to 2012, he was a Master Inventor and Research Staff Member at IBM T. J. Watson Research Center, New York, and has since held faculty positions in India and the US. His research focuses on brain-inspired computing, neuromorphic systems, and hardware accelerators. He has co-authored 100+ peer-reviewed papers, one monograph, an edited book, and holds 59 U.S. patents.  
\end{IEEEbiography}

\end{document}